\documentclass{aa}  
\usepackage{graphicx}
\usepackage{txfonts}
\usepackage{longtable}
\usepackage{pdflscape}
\usepackage{pdfpages}

\begin{document}

   \title{Searching for Magnetic White Dwarfs in LAMOST DR10}


   \author{Si-Cheng Yu 
      \inst{1,}
      \thanks{Corresponding author: sicheng.2.yu@oulu.fi}
   \and Juan-Juan Ren
      \inst{2, 3}
   \and Vitaly V. Neustroev
      \inst{1}
   \and Thomas Hackman
      \inst{4}      
   \and Hao-Tong Zhang
      \inst{3}
   \and Yi-Qiao Dong
      \inst{3}
   \and Zhong-Rui Bai
      \inst{3}
   \and Hai-Long Yuan
      \inst{3}
   \and Mengxin Wang
      \inst{3}
   \and Ming Zhou
      \inst{3}
   }

   \institute{Space Physics and Astronomy research unit, PO Box 3000, FI-90014 University of Oulu, Finland\\ \email{sicheng.2.yu@oulu.fi}
        \and
             Key Laboratory of Space Astronomy and Technology, National Astronomical Observatories, Chinese Academy of Sciences, Beijing 100101, China
        \and
             Key Laboratory of Optical Astronomy, National Astronomical Observatories, Chinese Academy of Sciences, Beijing 100101, China\\ \email{yuanhl@bao.ac.cn}
        \and
             Department of Physics, PO Box 64, 00014 University of Helsinki, Helsinki, Finland
    }
   \date{Received~~20xx month day; accepted~~20xx~~month day}

 
  \abstract
  {}
   {Magnetic white dwarfs (MWDs) are key to understanding the origin and evolution of magnetic fields in compact stars. While large spectroscopic surveys such as SDSS have greatly expanded the known sample, the potential of LAMOST has not yet been fully explored. Our aim is to identify and characterize isolated MWDs in the LAMOST DR10 database. }
   {We cross-matched LAMOST DR10 spectra with white dwarf candidates from \textit{Gaia} EDR3 and with recent SDSS-based catalogs of MWDs. Zeeman splitting in Balmer and helium absorption lines was used as the primary diagnostic to identify magnetic fields and to estimate their strengths. Reference objects from SDSS catalogs were used to test the detectability of MWDs in LAMOST low-resolution spectra. }
   {We identified 63 isolated MWDs in LAMOST DR10, of which 32 are new discoveries. Surface magnetic field strengths were measured from Zeeman splitting, covering a range from a few MG up to several tens of MG. For previously known SDSS MWDs, our LAMOST-based field measurements show mostly agreement with published values.  }
   {This work demonstrates the capability of LAMOST low-resolution spectroscopy to identify and characterize isolated MWDs. The newly discovered objects expand the known population and provide valuable targets for future high-resolution spectroscopic and polarimetric follow-up studies. Our results highlight the potential of combining LAMOST with \textit{Gaia} and other large surveys to build a more complete census of MWDs.  }
 

   \maketitle
%

\section{Introduction}
\label{sect:intro}
Magnetic white dwarfs (MWDs) represent a fascinating subset of white dwarf (WD) stars. Their magnetic field strengths span a broad range, from $10^4$ G to above $10^9$ G \citep{ferrario2015}, and they are typically detected via Zeeman splitting of spectral lines, cyclotron radiations or, in some cases, polarization signatures. Isolated MWDs are thought to originate through a range of evolutionary pathways, including single-star evolution \citep{Angel1985}, binary interactions during common-envelope evolution \citep{tout2008, schreiber2021}, or WD–WD mergers \citep{Berro2012}. A long-standing question is whether the magnetic field is a fossil remnant from the progenitor star (e.g. Ap or Bp type) or is generated/amplified during late stages of stellar evolution.

Statistically, $\sim$10--20\% of all WDs are magnetic \citep{kepler2013, ferrario2015}, though this fraction may be underestimated due to selection biases. Isolated MWDs tend to be more massive than their non-magnetic counterparts \citep{shaw2020}, suggesting either a merger origin or more massive progenitors. Population synthesis studies have sought to reconcile the observed mass distribution and field strengths with these formation channels, but no consensus has yet been reached.

The catalog of MWDs has expanded significantly in recent years, primarily thanks to large-scale spectroscopic surveys. The Sloan Digital Sky Survey (SDSS; \citealt{york2000}) has played a particularly important role, leading to an exponential increase in the number of known MWDs \citep{gansicke2002, schmidt2003, vanlandingham2005, kulebi2009, kepler2013, ferrario2015}. The most recent SDSS-based catalogs report over 800 MWDs, including both hydrogen-rich (DAH) and helium-rich (DBH) types \citep{kepler2013, Amorim2023, Hardy2023}, enabling detailed statistical studies of their properties.

In this work, we conducted a systematic search for MWDs using the low-resolution spectroscopic (LRS) database provided by the Large Sky Area Multi-Object Fiber Spectroscopic Telescope (LAMOST) DR10. By cross-matching with existing WD catalogs and employing Zeeman splitting as the main diagnostic, we identified and measured the surface magnetic field strengths of 63 isolated MWDs having 78 spectra in the LAMOST database, including 38 spectra of 32 newly discovered objects. This sample both complements and extends previous SDSS-based studies, and demonstrates the capability of LAMOST to contribute significantly to the census of MWDs.

\section{LAMOST Spectroscopy}
The LAMOST, also known as the Guo Shoujing Telescope, is a quasi-meridian reflecting Schmidt telescope with an effective aperture of approximately 4 meters and a wide field of view of 5 degrees \citep{wang1996, su2004, cui2012}. Located at the Xinglong Station of the National Astronomical Observatories, Chinese Academy of Sciences, LAMOST is equipped with 4000 optical fibers distributed across its focal plane. Each fiber is connected to one of 16 spectrographs, enabling the simultaneous acquisition of over 4000 spectra per exposure. Each spectrograph includes two CCD cameras, covering both blue and red wavelength channels. The LAMOST spectral range spans from 3700{\AA }\ to 9000{\AA }, with a resolving power of $R \sim 1800$.
The raw data collected during the survey are processed using the LAMOST two-dimensional (2D) pipeline \citep{bai2017RAA, bai2017PASP}. This pipeline performs dark and bias subtraction, cosmic ray removal, one-dimensional spectral extraction, co-addition of sub-exposures (each plate typically being observed at least three times), and stitching of the blue and red channel spectra. The resulting combined spectra are then passed through the LAMOST one-dimensional (1D) pipeline, which performs spectral classification and redshift measurement. 

LAMOST Data Release 10 (DR10) comprises a total of 11,441,011 spectra from 5923 observed plates. Of these, 11,100,139 are classified as stellar spectra, from which our initial sample groups are drawn.
It employs a universal target selection strategy, providing more uniform coverage across stellar populations, and has proven effective in identifying numerous new WD–main sequence (WDMS) binaries \citep{ren2013, ren2014, ren2018}. 
Given its wide-area sky coverage and unique survey design, we intend to believe that it has the great potential to provide a complementary and underutilized resource for identifying MWDs. 

\section{Identification of MWDs in LAMOST DR10}

Our sample was constructed by cross-matching LAMOST DR10 with the catalogue of WD candidates from \textit{Gaia} EDR3 \citep{GentileFusillo2021}. We located 10,279 LAMOST spectra within 3 arcsec of \textit{Gaia} WDs (and WD candidates).\footnote{This preliminary list of WDs can be found in electronic supplementary material.}
We also cross-matched LAMOST DR10 with the latest SDSS-based catalogues of DAH (hydrogen-atmosphere) and DBH (helium-atmosphere) MWDs compiled by \citet{Amorim2023} and \citet{Hardy2023}, respectively. This provided a reference set of confirmed MWDs that helps with finding more targets with similar features; and allowed us to assess the effectiveness of LAMOST spectra in recovering them.  

Zeeman splitting was the primary diagnostic for identifying magnetic fields in WDs. This is a direct and robust indicator, as most WDs exhibit strong and clean hydrogen (or helium) absorption lines in the optical regime. The splitting of these lines provides direct evidence of the presence of a magnetic field, and, when the features are clearly resolved, also enables an estimate of the field strength.

For MWDs exhibiting Zeeman-split spectral lines, the mean surface magnetic field can be estimated from the separation of split components.  
In the weak-field regime ($B \lesssim 2$~MG), lower-level ($n\leq3$) Balmer lines are split into three components, with a separation given by:
\begin{equation}\label{eq:zeeman_simple}
\Delta \lambda = \pm 4.67 \times 10^{-7} \lambda^2 B \quad (\text{\AA})
\end{equation}
where $\lambda$ is in \AA\ and $B$ is in MG.\footnote{For completeness, the full quadratic Zeeman effect for higher fields is given by
\begin{equation*}
\Delta\lambda_q \approx -4.97 \times 10^{-23} \lambda^2 n^4 (1+m_l^2) B^2 \, (\text{\AA}),
\end{equation*}
where $n$ is the principal quantum number and $m_l$ the magnetic quantum number.}

\begin{figure}
\centering
\includegraphics[width=1.0\linewidth]{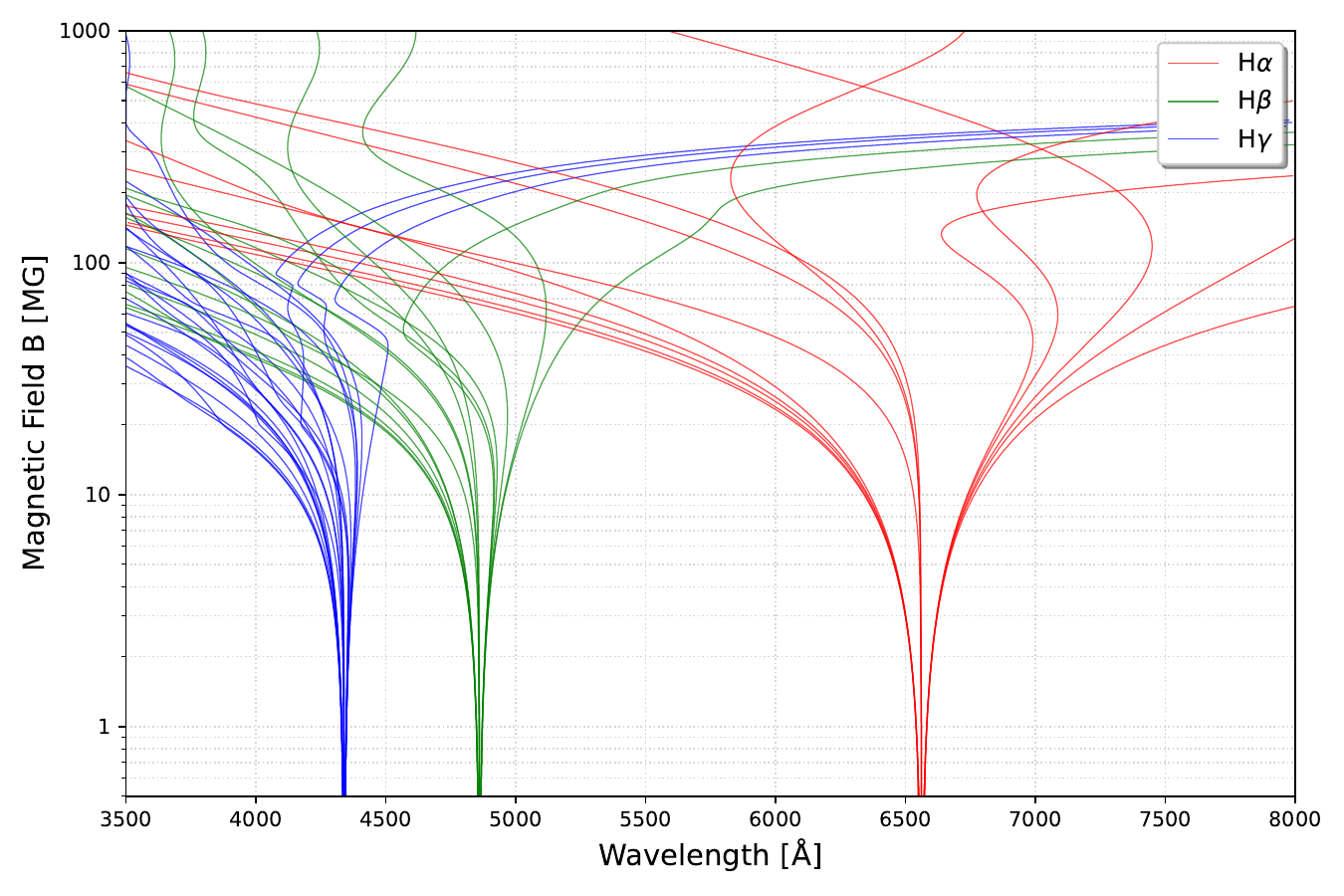}
\caption{Theoretical Zeeman Splitting distribution as a function of magnetic field strength for the hydrogen Balmer lines H$\alpha$, H$\beta$, and H$\gamma$ transitions calculated based on the numerical models of \citet{Schimeczek2014}.}
\label{fig:Schimeczek}
\end{figure}

However, for higher strength fields, the linear Zeeman approximation becomes inadequate as the magnetic and Coulomb forces reach comparable magnitudes. For this reason, we decided to utilize the high-precision numerical framework provided by \citet{Schimeczek2014}, which solves the non-relativistic Schrödinger equation for the hydrogen atom across the full range of field strengths ($0$ to $4.7 \times 10^3$~MG) relevant to MWDs. Their approach employs an expansion in two-dimensional B-splines to determine energy levels and dipole strengths for states up to $n=15$. This method captures the complex $l$-mixing and the behavior of energy levels near anticrossings, where transitions can shift rapidly. The resulting relationship between magnetic field strength and transition wavelengths for the H$\alpha$, H$\beta$, and H$\gamma$ series is illustrated in Figure~\ref{fig:Schimeczek}.\footnote{The atomic transition data calculated by \citet{Schimeczek2014} is no longer available through the website provided in their original article, but can be found at \\ \url{https://doi.org/10.18419/DARUS-2118}.}

\begin{figure}
\centering
\includegraphics[width=1.0\linewidth]{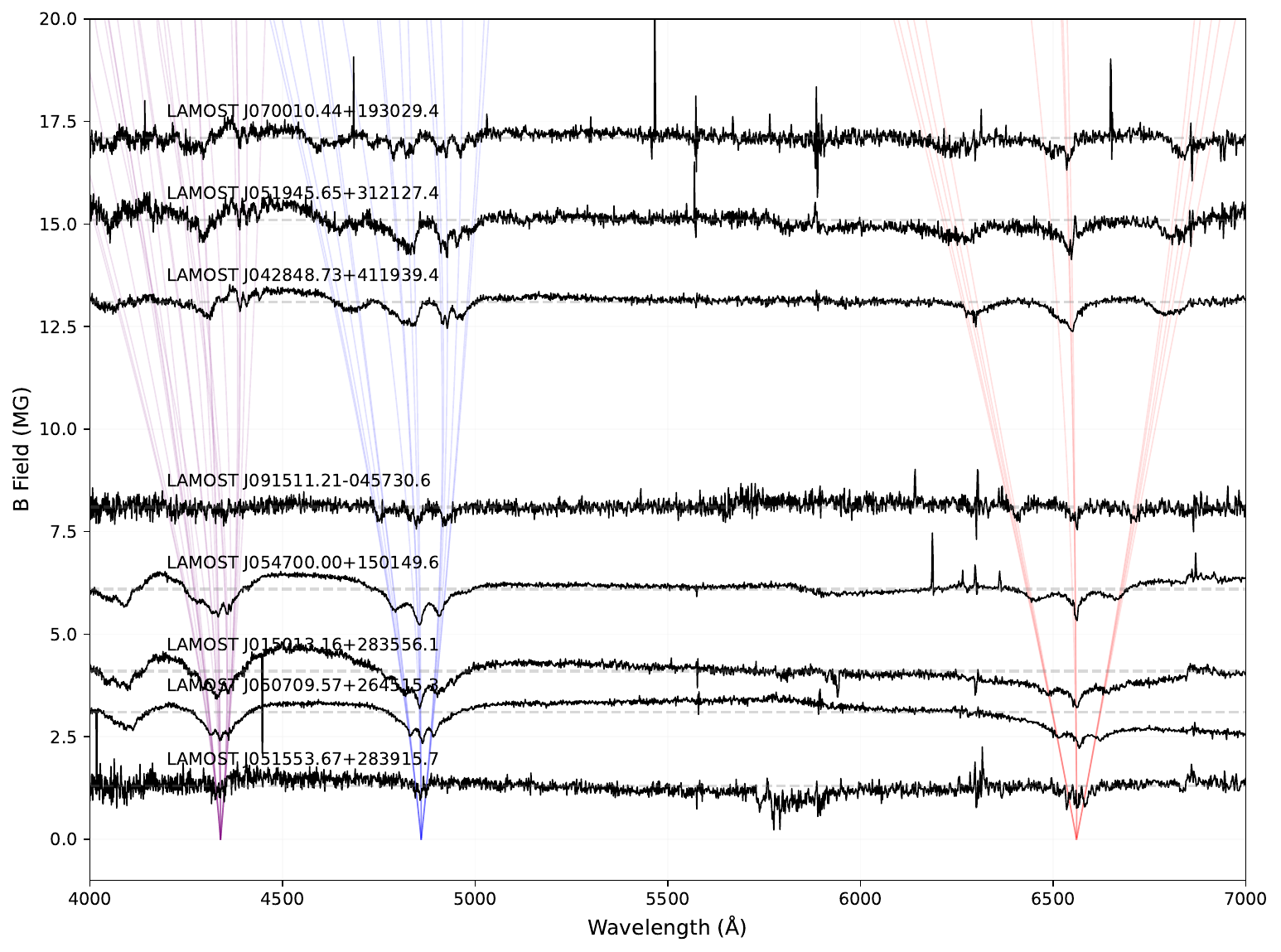}
\caption{Several LAMOST isolated MWD spectra with the Zeeman effect on their H$\alpha$, H$\beta$ and H$\gamma$ lines in the linear regimes. spectra were normalized to make better comparisons. As can be seen, the strength of derived magnetic fields are positively correlated to the separation among sub-components of Balmer lines caused by Zeeman-Splitting.}
\label{fig:mwdzeeman_lamost}
\end{figure}

In our analysis, we compared the LAMOST DR10 spectra against these theoretical stationary transitions for the Balmer series. By shifting the observed spectra to the rest frame and comparing the morphology of H$\alpha$, H$\beta$, and H$\gamma$ to the Schimeczek transitions, we were able to provide robust identifications and estimate the mean surface magnetic field $B$. This method allows for the detection of fields even when the individual Zeeman components are not fully resolved, as the overall profile shape remains distinct. We confirmed the magnetic nature of stars previously identified in the SDSS-based catalogues of \citet{Amorim2023} and successfully identified new MWD candidates. A series of examples in diverse field strengths are shown in Figure~\ref{fig:mwdzeeman_lamost}. 

The resulting catalogue includes the LAMOST designation, \textit{Gaia} DR3 source ID, and the best-fit mean surface magnetic field $B$. Our results show a broad distribution of field strengths ranging from the detection limit of $\sim 0.5$~MG up to approximately $15$~MG, demonstrating the capability of LAMOST's medium-to-low resolution spectroscopy in characterizing the magnetic properties of the local white dwarf population.

The relation between the observed mean field strength and the polar field strength, assuming a dipolar surface field, is:
\begin{equation}\label{eq:BpBo}
B = \frac{1}{2} B_p \sqrt{1 + 3\cos^2\theta}
\end{equation}
where $B_p$ is the polar field and $\theta$ is the angle between the field and the line of sight. Due to limitation in spectral resolution and diverse qualities of our spectra, this work only estimates the mean field strength and does not take into account the line shapes. Our estimation also ignores binarity or any effect beyond dipole moments. This approach is different from that used by \citet{Amorim2023} to calculate the field strengths and can explain occasional inconsistency between our and their measurements as they reported the $B_{\rm p}$ values.

With the aid of reference targets and field estimations from Zeeman splitting, we visually inspected LAMOST WD spectra and identified 85 isolated MWDs, including 31 newly discovered objects, for which we also estimated their surface magnetic fields. 

\section{Results}
\label{sect:Result}

A total of 129 common targets were found by cross-matching the LAMOST DR10 low-resolution spectroscopic database with the SDSS MWD catalogues \citep{kepler2013, Amorim2023, Hardy2023}, using a matching radius of 2 arcseconds. Visual inspection of these matched spectra revealed 31 objects exhibiting significant Zeeman splitting features. Using these SDSS-confirmed MWDs as reference examples, we extended our search to a broader sample of 10,279 WD candidates obtained by cross-matching LAMOST DR10 with the \textit{Gaia} EDR3 WD catalogue \citep{GentileFusillo2021}. This effort resulted in the identification of 63 LAMOST spectra displaying clear magnetic features—representing newly discovered MWDs in the LAMOST dataset (some of which include multiple exposures of the same object).

The full list of isolated MWDs identified in LAMOST DR10, including both previously known and newly discovered systems, is presented in the appendix~\ref{appendix:iso_mwd_list}. Their spectra are shown in the appendix~\ref{appendix:spectra}.

\subsection{LAMOST J1538+0842}
\begin{figure}
	\begin{center}
		\includegraphics[width=1.0 \linewidth]{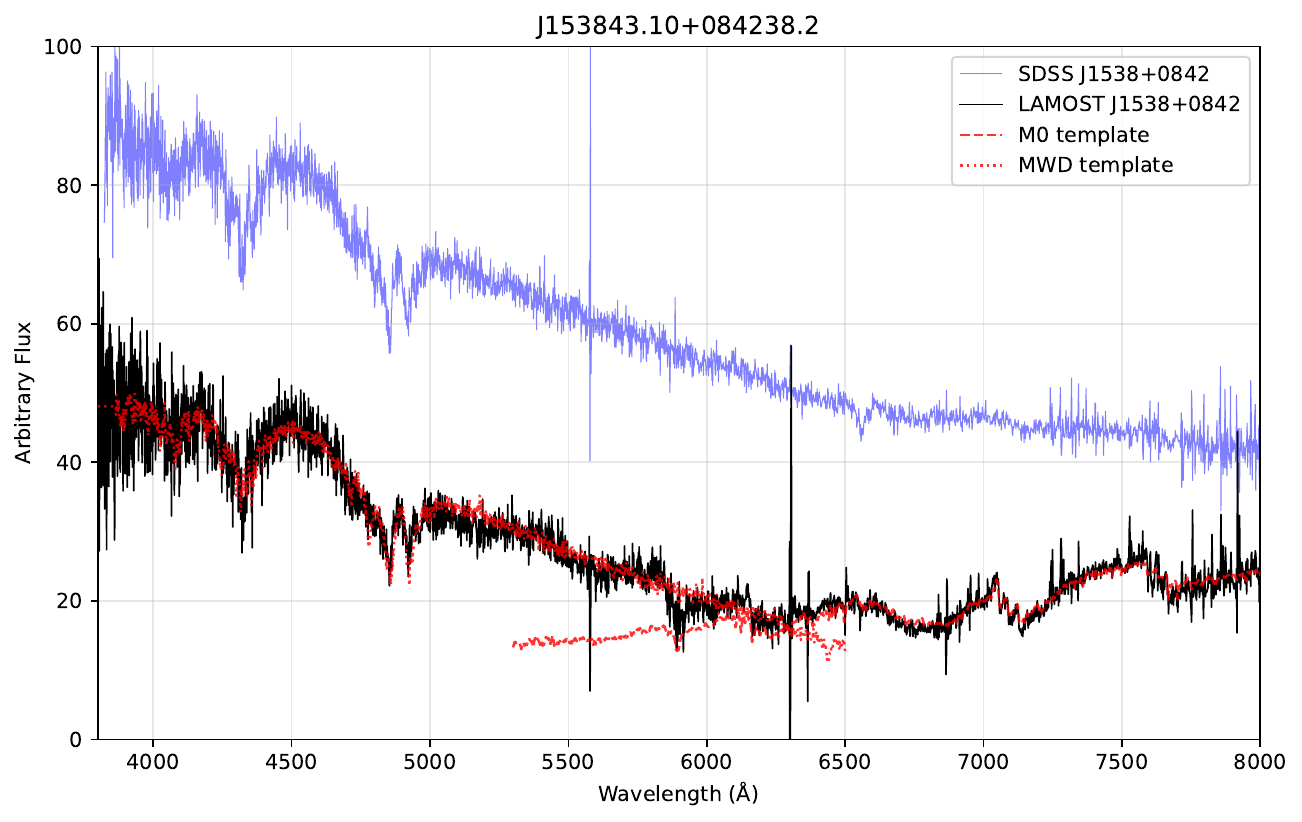}
		\caption{\label{fig:J1538+0842} 
		Composite spectral analysis of J1538+0842. The LAMOST spectrum (black) is compared with an MWD template with a surface field of $B \approx 12$\,MG (red dotted line) and an M0 star template (red dashed line). While the MWD template fits the Zeeman-split Balmer lines in the blue ($\lambda < 6500$\,\AA), the M0 template is required to reproduce the molecular bands in the red ($\lambda > 5500$\,\AA). On the other hand, the SDSS spectrum of the same target (blue), scaled for comparison, shows no significant M star features in the red.}
	\end{center} 
\end{figure}

Among the identified MWDs, J1538+0842 (\object{SDSS J153843.10+084238.2}) presents an interesting case. The LAMOST spectrum (Figure~\ref{fig:J1538+0842}) exhibits the clear Zeeman-split Balmer lines of an MWD with a surface field of $\sim$12\,MG, but is superposed with prominent TiO bands and molecular features characteristic of an M-type main-sequence (MS) star at the red end ($\lambda > 5500$\,\AA). However, the latter features are not seen in the SDSS spectrum of this object.

\begin{figure}
	\begin{center}
		\includegraphics[width=0.8 \linewidth]{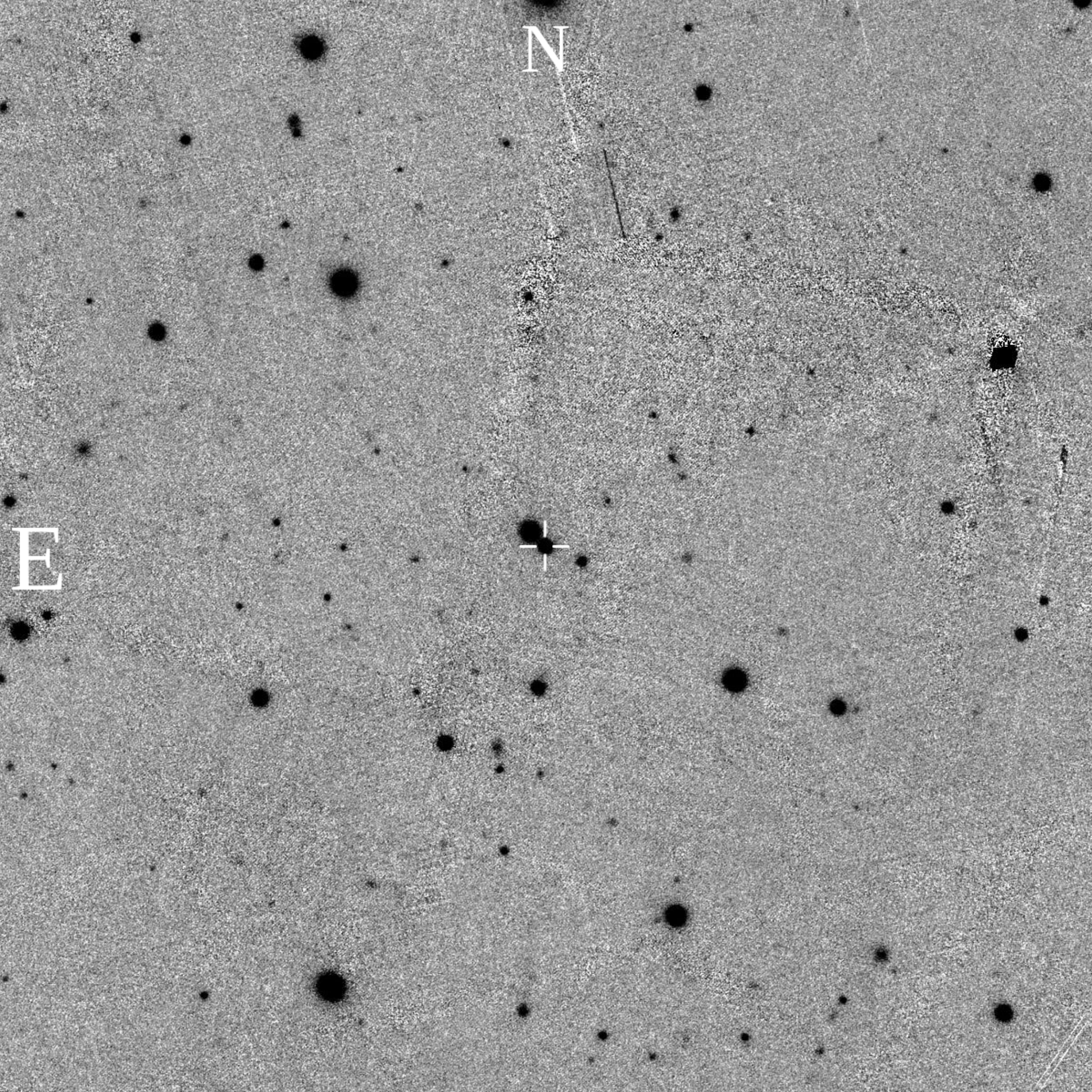}
		\caption{\label{fig:1538+0842OB} 
		The 6'$\times$6' field-of-view of J1538+0842, the star in the cursor is our target MWD, a M type main sequence star is sitting roughly 7 arcsec away from it. }   
	\end{center}
\end{figure}

Analysis of the sky atlas (Figure~\ref{fig:1538+0842OB}) reveals an M-star companion, 2.6 magnitude brighter than our target in the SDSS $r'$ band, located $\sim$6.14$\arcsec$ from the MWD. Given the LAMOST fiber radius of $\sim$3.3$\arcsec$, and that the spectrum was taken under non-optimal weather conditions, we believe that the MS features in the spectrum are not intrinsic to the MWD's immediate environment but are the result of flux contamination from this neighbor. 

However, we point out that \textit{Gaia} DR3 astrometry strongly suggests that this proximity is not a chance alignment. Both stars share nearly identical parallaxes ($11.81 \pm 0.12$\,mas and $11.80 \pm 0.02$\,mas) and consistent proper motions: for the MWD, $\mu_{\alpha}\cos\delta = 15.52 \pm 0.12$\,mas\,yr$^{-1}$, $\mu_{\delta} = -65.16 \pm 0.11$\,mas\,yr$^{-1}$; for the M dwarf, $\mu_{\alpha}\cos\delta = 14.73 \pm 0.02$\,mas\,yr$^{-1}$, $\mu_{\delta} = -65.86 \pm 0.02$\,mas\,yr$^{-1}$.
Adopting a common parallax of $11.80$\,mas, the angular separation of 6.14$\arcsec$ corresponds to a minimum projected physical separation of $\sim$520\,AU. While the current data cannot definitively confirm a bound orbit, the matching astrometric parameters suggest that J1538+0842 and its neighbor may constitute a wide, fully detached MWD+MS binary system.

Such systems have long been predicted by the classical fossil field theory, which states that the magnetic fields in MWDs are inherited from their Ap or Bp progenitors \citep{Angel1981}. Under this theory, if MWDs primarily result from single-star evolution, their presence in detached binaries should be as significant as their occurrence as isolated stars or in close interacting binaries. However, detached MWD+MS binary systems have been notably absent from observational records \citep{Liebert2005}. Such "missing" population has driven the development of alternative scenarios that favor binary interactions as the primary mechanism for magnetic field generation \citep{tout2008, wickramasinghe2014, belloni2020}. Thus, J1538+0842 represents a high-potential candidate for the first fully detached MWD+MS binary.

\section{Discussion}

\subsection{Comparison Between LAMOST and SDSS MWD Catalog}
\begin{figure}
	\begin{center}
		\includegraphics[width=1.0 \linewidth]{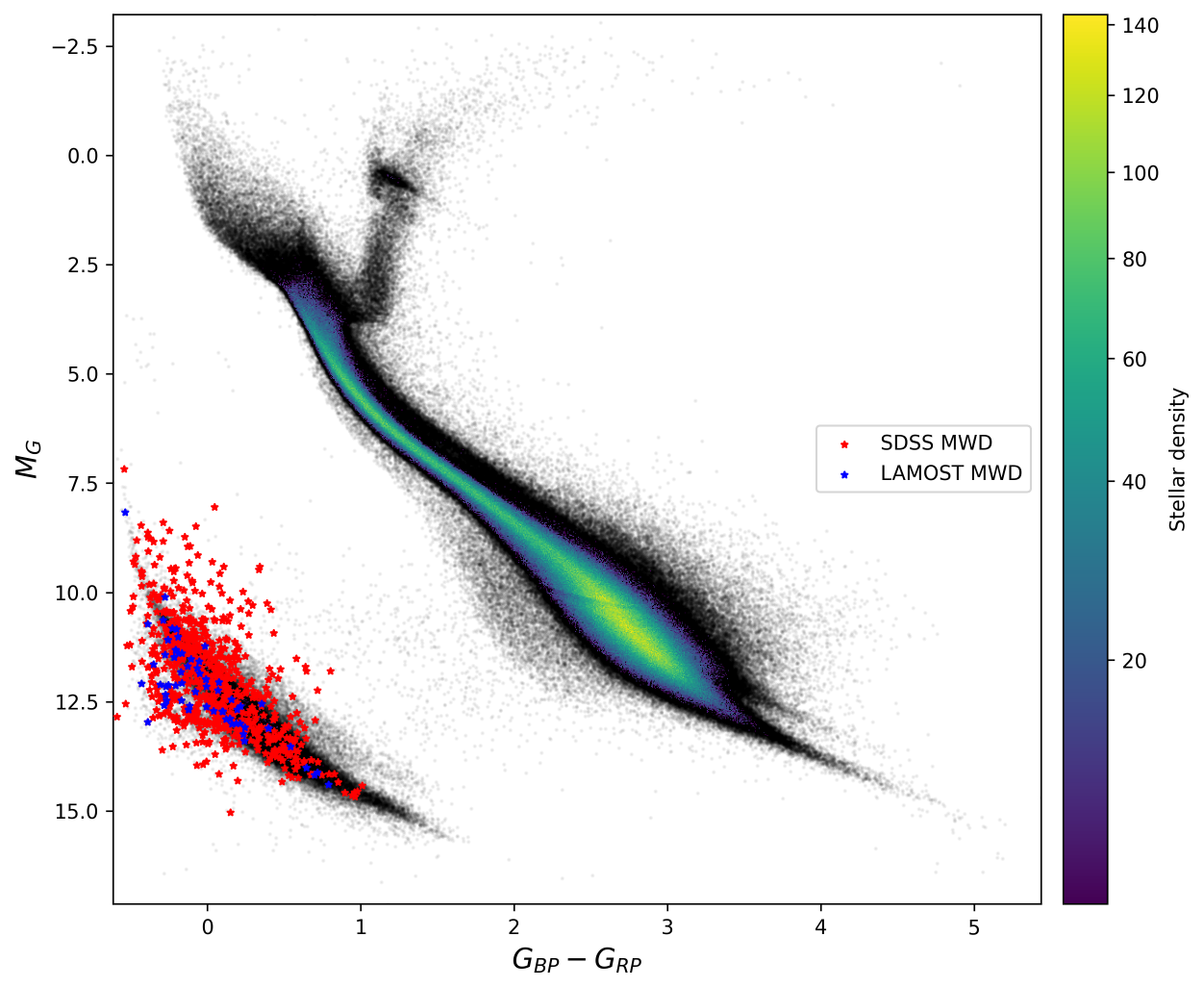}
		\caption{\label{fig:hr_mwd} 
        Hertzsprung–Russell diagram showing the locations of MWDs.
        Red stars mark SDSS MWDs and blue stars mark LAMOST MWDs. 
        Both groups are distributed uniformly along the WD cooling sequence.}
    \end{center} 
\end{figure}

Initially, we expected LAMOST to recover relatively cooler MWDs, under the assumption that SDSS preferentially selected hotter and bluer objects, since it was primarily designed to target quasars and galaxies, a population that overlaps with hot WDs in color space \citep{richards2002, Smolcic2004}. 
However, both LAMOST and SDSS MWDs are similarly distributed in the WD branch seen in the Hertzsprung–Russell diagram (Figure~\ref{fig:hr_mwd}). 
In detail, the SDSS MWD catalog, comprising more than 800 objects, reveals an average effective temperature of $13{,}770 \pm 5884$~K and an average surface gravity of $\log g = 8.17 \pm 0.53$ \citep{Amorim2023, Hardy2023}. In comparison, our LAMOST MWD catalog contains 63 targets, with an average effective temperature of $14{,}548 \pm 5697$~K and an average surface gravity of $\log g = 8.38 \pm 0.36$. Differences in effective temperature and surface gravity are statistically insignificant within the quoted dispersions.
This outcome suggests that the limited depth and spectral quality of LAMOST observations may have hindered the detection of cooler MWDs. Consequently, the selection effects of SDSS with respect to MWD temperature appear to be less significant than originally anticipated.

\subsection{On the Origin of Magnetic Fields in MWDs}
The origin of strong MWDs remains controversial, as classical theories often conflict with observational results. Historically, MWD magnetic fields were assumed to be fossil remnants from magnetic Ap and Bp stars, preserved through stellar evolution via magnetic flux conservation \citep{Angel1981}. However, the rapid increase in confirmed MWDs \citep{gansicke2002, schmidt2003, vanlandingham2005} has highlighted inconsistencies between theoretical predictions and observations.

For example, population synthesis studies of SDSS MWDs by \citet{Liebert2005} challenged the fossil field model by noting the near absence of fully detached MWD-main-sequence binaries. If MWD progenitors evolved independently, such detached binaries should be relatively common. Furthermore, \citet{kawka2007} showed that the birth rates of Ap and Bp stars are insufficient to account for the observed fraction of MWDs among evolved stars. These discrepancies indicate that classical single-star scenarios alone cannot fully explain MWD magnetism.

Several alternative formation scenarios have been proposed. The coalescing double WD scenario predicts magnetic field generation in the corona of the merger remnant \citep{Berro2012}, while the crystallization scenario suggests that cooling and solidification of the WD interior can amplify pre-existing fields \citep{isern2017}. Both scenarios have explanatory power for some MWDs but cannot account for the full range of observed field strengths.

A more widely favored hypothesis links high magnetic fields to binary interactions. \citet{tout2008} proposed that MWD magnetism originates during the common-envelope (CE) phase, with closer binary interactions producing stronger fields. This idea was further supported by \citet{wickramasinghe2014}, who developed a dynamo model in which differential rotation during binary interactions amplifies toroidal fields inside the WD. 

\citet{schreiber2021} recently proposed a unified dynamo framework in which both internal and binary-driven processes contribute to the formation of MWDs. In this model, convective motions induced by core crystallization in cooling WDs can sustain a dynamo that generates moderate magnetic fields; with the influence from a donor star, accretion-driven spin-up and tidal effects amplify differential rotation and trigger much stronger dynamos. These two channels are not mutually exclusive but act together to explain the wide variety of magnetic field strengths and current MWD configurations. 

While the dual-channel dynamo framework provides an attractive explanation for the overall diversity of MWDs, our discovery of a fully detached MWD+MS binary does not fit naturally into this picture. The presence of a strong-field MWD in a detached binary system raises questions: unless its magnetism is the relic of a fossil field from an Ap or Bp progenitor, the most plausible alternative is that this system originated in a hierarchical triple, with the present-day MWD formed through the merger of an inner companion. The relatively high mass of the WD ($0.875\,M_\odot$; \citealt{Kilic2025}), its $\log g=8.44$, and effective temperature of $\sim9000$\,K further strengthen the case for a merger origin. 

Our discovery of a fully detached MWD+MS system, LAMOST J1538+0842, provides a critical observational counterpoint. The presence of a strong magnetic field in a detached configuration demonstrates that intense magnetism can exist without ongoing mass transfer or immediate post-CE interaction. This finding supports the idea that internal dynamo mechanisms, potentially combined with earlier binary interactions or crystallization processes, may contribute to magnetic field generation. Furthermore, it suggests that detached magnetic systems, though rare, represent an important evolutionary stage that should be incorporated into models of MWD formation and evolution.

\section{Conclusions}
\label{sect:conclusion}

In this work, we have presented the first systematic search for isolated MWDs in the LAMOST DR10 spectroscopic database. We identified a total of 78 LAMOST spectra of 63 isolated MWDs, including 38 spectra of 32 newly discovered objects. Our analysis indicates that LAMOST is capable of detecting MWDs across a range of temperatures and surface gravities, complementary to SDSS and with the potential to mitigate selection biases inherent to previous surveys.
Comparisons between the LAMOST and SDSS MWD populations show no statistically significant differences. This indicates that the current selection effect on MWDs is negligible. Nevertheless, given its wide-area sky coverage and balanced sampling of stellar populations, LAMOST represents a promising resource for further expanding the known population of MWDs. 
Overall, our results highlight LAMOST’s capability to contribute to the study of isolated MWDs, providing new targets for follow-up spectroscopic and photometric investigations, and helping to improve our understanding of the formation and evolution of magnetic fields in WDs.

\begin{acknowledgements}
We would like to thank the anonymous referee for useful comments and suggestions.
We thank Dr. Schimeczek for kindly sharing his detailed calculation of atomic data for the spectral analysis of DA MWDs, which is crucial to our work. 
Sicheng Yu is funded by the Vilho, Yrjö and Kalle Väisälä Foundation of the Finnish Academy of Science and Letters. 
Haotong Zhang thanks the support of the National Natural Science Foundation of China (12273056) and the National Key R\&D Program of China (Grant No. 2022YFA1603002). 
Juanjuan Ren thanks the support of the China Manned Space Program (Grant No. CMS-CSST-2025-A19).
Hailong Yuan and Zongrui Bai are supported by the National Key R\&D Program of China (Grant No. 2023YFA1607901). 
Ming Zhou is supported by the National Natural Science for Youth Foundation of China (Grant No. 12503091)
This work is supported by and based on the survey of Guoshoujing Telescope. Guoshoujing Telescope (the Large Sky Area Multi-Object Fiber Spectroscopic Telescope, LAMOST) is a National Major Scientific Project built by the Chinese Academy of Sciences. 

\end{acknowledgements}

\bibliographystyle{aa}

\begin{appendix}

\appendix
\onecolumn
\section{Full Catalogue of Isolated MWDs}
\label{appendix:iso_mwd_list}
The following table lists all MWDs identified in this work. The columns in the table are structured as follows:

\begin{enumerate}
    \item \textbf{Designation:} Target designation from celestial coordinates (RA and Dec, in degrees). For example, J015013.17+283556.2 corresponds to $RA = 01^\mathrm{h}50^\mathrm{m}13.17^\mathrm{s}$, $Dec = +28^\circ35'56.2''$.
    
    \item \textbf{Gaia ID:} Target ID number from \textit{Gaia} EDR3 catalog.  
    
    \item \textbf{WD type:} Indicates whether the target MWD is hydrogen-rich (DA), helium-rich (DB) or carbon-rich (DQ). 

    \item \textbf{$B_{\rm LAMOST}$:} Mean magnetic field estimated with LAMOST spectra in MG.
    
    \item \textbf{$B_{\rm SDSS}$}: Magnetic field strength from SDSS MWD catalogues \citep{Amorim2023, Hardy2023}.
                
    \item \textbf{Plx:} Parallax (mas) from \textit{Gaia} EDR3.
    
    \item \textbf{$BP{-}RP$:} $BP{-}RP$ colors from \textit{Gaia} EDR3.
        
    \item \textbf{$G$:} $G$ magnitudes from \textit{Gaia} EDR3.
    
    \item \textbf{$T_{\rm eff}$:} Effective temperature from fitting dereddened $G$, $BP$, and $RP$ absolute fluxes with pure-H model atmospheres, from the \textit{Gaia} EDR3 WD catalogue \citep{GentileFusillo2021}, with the caveat that the models do not include magnetic field.
    
    \item \textbf{$\log g$:} Surface gravity from fitting dereddened $G$, $BP$, and $RP$ absolute fluxes with pure-H model atmospheres, from the \textit{Gaia} EDR3 WD catalogue \citep{GentileFusillo2021}, with the caveat that the models do not include magnetic field.
\end{enumerate}

\noindent
The full table of identified MWDs, including all measured Zeeman splits and derived magnetic fields, is provided below.

\begin{longtable}{llllllllll}
\caption{List of LAMOST isolate MWDs with observational details and basic parameters.} 
\label{tab:iso_MWD_list}\\

\hline\hline
Designation & \textit{Gaia} ID & WD type & $B_{\rm LAMOST}$ & $B_{\rm SDSS}$ & $\mathrm{Plx}$ & $BP{-}RP$ & $G$ & $T_{\rm eff}$ & $\log g$ \\
 & & & (MG) & (MG) & (mas) & (mag) & (mag) & (K) & (cgs) \\
\hline
\endfirsthead

\caption{continued. List of isolated MWDs in LAMOST DR10.} \\
\hline

\hline
\endhead

\hline
\endfoot

J001742.43+004137.4 & 2545579360599665792 & DB & - & 8.17 & 7.88 & -0.11 & 17.05 & 14252 & 8.06 \\
J003634.76+111808.4 & 2751415904481932416 & DA & 1.5 & - & 5.77 & -0.17 & 17.59 & 16039 & 8.11 \\
J015013.16+283556.1 & 299265624604662656 & DA & 4.5 & - & 13.93 & -0.01 & 16.89 & 12364 & 8.60 \\
J015013.17+283556.1 & 299265624604662656 & DA & 4.1 & - & 13.93 & -0.01 & 16.89 & 12364 & 8.60 \\
J024822.73+160013.4 & 33656531962611072 & DA & 1 & - & 12.17 & -0.25 & 16.69 & 20299 & 8.78 \\
J030407.39-002541.9 & 3266337781252698880 & DA & 6.3 & 10.48 & 7.36 & -0.01 & 17.82 & 12527 & 8.34 \\
J033253.91+284006.9 & 119677790531181056 & DA & 6.4 & - & 7.33 & -0.05 & 17.25 & 13718 & 8.01 \\
J041522.02+374746.2 & 225787530263448576 & DA & 5.9 & - & 10.33 & 0.21 & 17.54 & 9637 & 8.18 \\
J042848.73+411939.4 & 180107598134288640 & DA & 13.4 & - & 8.70 & -0.07 & 17.03 & 13699 & 8.13 \\
J044350.77+265124.7 & 154537772330054656 & DA & 1.2 & - & 23.24 & 0.79 & 17.57 & 5895 & 8.20 \\
J050259.44+240248.3 & 3418741367154043648 & DA & 1.7 & - & 12.18 & 0.15 & 17.46 & 10210 & 8.47 \\
J050259.44+240248.3 & 3418741367154043648 & DA & 1.8 & - & 12.18 & 0.15 & 17.46 & 10210 & 8.47 \\
J050709.57+264515.3 & 3421894079307215744 & DA & 2.7 & - & 18.62 & -0.28 & 16.23 & 21692 & 9.07 \\
J050709.57+264515.3 & 3421894079307215744 & DA & 2.7 & - & 18.62 & -0.28 & 16.23 & 21692 & 9.07 \\
J051553.67+283915.7 & 3422405214775411840 & DA & 1.3 & - & 33.26 & 0.65 & 16.40 & 6411 & 8.18 \\
J051553.67+283915.7 & 3422405214775411840 & DA & 1.4 & - & 33.26 & 0.65 & 16.40 & 6411 & 8.18 \\
J051945.65+312127.3 & 180333655149488512 & DA & 14.5 & - & 8.77 & -0.01 & 17.27 & 12556 & 8.22 \\
J051945.65+312127.4 & 180333655149488512 & DA & 14.2 & - & 8.77 & -0.01 & 17.27 & 12556 & 8.22 \\
J051945.65+312127.4 & 180333655149488512 & DA & 15.1 & - & 8.77 & -0.01 & 17.27 & 12556 & 8.22 \\
J054700.00+150149.6 & 3347953532952671360 & DA & 5.4 & - & 14.51 & -0.17 & 16.27 & 16401 & 8.56 \\
J054959.99+182044.3 & 3398208399823387392 & DA & 4 & - & 6.76 & -0.22 & 17.33 & 18272 & 8.31 \\
J054959.99+182044.3 & 3398208399823387392 & DA & 2.9 & - & 6.76 & -0.22 & 17.33 & 18272 & 8.31 \\
J055119.55-001018.6 & 3218697767783768320 & DQ & - & - & 89.11 & 0.70 & 14.43 & 6186 & 8.20 \\
J060755.70+341525.9 & 3452373568124842752 & DA & 2 & - & 13.51 & -0.27 & 16.68 & 21359 & 8.93 \\
J060755.70+341525.9 & 3452373568124842752 & DA & 2.20: & - & 13.51 & -0.27 & 16.68 & 21359 & 8.93 \\
J063235.85+555902.1 & 995112350178946048 & DA & 1.1 & - & 27.00 & 0.16 & 15.86 & 9996 & 8.52 \\
J064058.16+151737.9 & 3356235604288817792 & DA & 1.2 & - & 9.23 & 0.07 & 17.23 & 10911 & 8.08 \\
J070010.44+193029.4 & 3365090177584636672 & DA & 14.3 & - & 10.21 & 0.04 & 17.67 & 11944 & 8.62 \\
J074213.46+315702.6 & 880354496226790400 & DA & - & 41.4 & 14.96 & 0.19 & 16.81 & 9590 & 8.23 \\
J083047.22+505734.2 & 1028123052204089472 & DA & 7.6 & - & 3.22 & -0.39 & 18.19 & 27857 & 8.34 \\
J085618.94+161103.6 & 611401999180118528 & DB & - & 1.5 & 13.87 & -0.21 & 15.60 & 17457 & 8.16 \\
J085618.94+161103.6 & 611401999180118528 & DB & - & 1.5 & 13.87 & -0.21 & 15.60 & 17457 & 8.16 \\
J085618.94+161103.6 & 611401999180118528 & DB & - & 1.5 & 13.87 & -0.21 & 15.60 & 17457 & 8.16 \\
J085618.94+161103.6 & 611401999180118528 & DB & - & 1.5 & 13.87 & -0.21 & 15.60 & 17457 & 8.16 \\
J091437.35+054453.3 & 580558842594128256 & DA & 5 & 9.23 & 7.35 & -0.17 & 17.49 & 16033 & 8.39 \\
J091511.21-045730.6 & 5758736375755925760 & DA & 8 & - & 2.26 & -0.54 & 16.39 & - & - \\
J093415.97+294500.4 & 696350817298141952 & DA & 14 & 32.01 & 2.90 & -0.28 & 19.11 & 18932 & 8.32 \\
J093447.89+503312.1 & 1018436320403748480 & DA & 3.9 & 6.61 & 7.39 & 0.24 & 18.74 & 9015 & 8.35 \\
J093903.33+114418.6 & 613742477182156800 & DA & 3.7 & - & 6.67 & -0.02 & 17.10 & 12097 & 7.66 \\
J094235.02+205208.3 & 639900580361748736 & DA & - & 62.79 & 5.11 & -0.27 & 18.61 & 20497 & 8.80 \\
J094719.41+484123.0 & 825180968385183616 & DA & 1.50: & - & 2.40 & -0.28 & 18.21 & 21295 & 7.62 \\
J100657.51+303338.0 & 745425074966126976 & DA & 3.7 & 3.23 & 5.33 & 0.16 & 18.82 & 10020 & 8.17 \\
J100715.55+123709.5 & 3881621915471688832 & DA & 3.3 & 5.08 & 3.60 & -0.13 & 18.91 & 16115 & 8.28 \\
J101428.09+365724.3 & 754143965951349760 & DA & 7.2 & 14.93 & 4.74 & 0.07 & 18.86 & 11015 & 8.21 \\
J103430.15+032736.1 & 3856950175919062144 & DA & 7.1 & 13.28 & 12.80 & -0.12 & 17.05 & 14685 & 8.75 \\
J105544.85+211104.4 & 3988212592756945152 & DA & 0.7 & 0.5 & 21.17 & 0.72 & 17.50 & 6136 & 8.14 \\
J110914.98+031808.6 & 3811719742181133696 & DA & 0.5 & - & 6.30 & -0.19 & 17.02 & 16599 & 7.90 \\
J111004.26+401853.3 & 765425535072834816 & DA & 1 & - & 3.35 & -0.20 & 18.22 & 15577 & 7.70 \\
J111812.67+095241.3 & 3915154645001054976 & DA & 2.4 & 2.27 & 6.05 & 0.09 & 18.82 & 11376 & 8.56 \\
J112257.10+322327.7 & 757328563942257152 & DA & 7.6 & 12.87 & 4.51 & -0.12 & 19.41 & 12347 & 8.63 \\
J112257.10+322327.7 & 757328563942257152 & DA & 6.8 & 12.87 & 4.51 & -0.12 & 19.41 & 12347 & 8.63 \\
J112257.10+322327.7 & 757328563942257152 & DA & 6.7 & 12.87 & 4.51 & -0.12 & 19.41 & 12347 & 8.63 \\
J121033.23+221403.0 & 4001466277717538432 & DA & 1.7 & 2.23 & 10.35 & -0.08 & 17.19 & 13879 & 8.51 \\
J121735.23+082810.1 & 3902183809407583872 & DA & 2.1 & 2.57 & 12.34 & -0.21 & 16.68 & 17476 & 8.65 \\
J122249.13+481133.1 & 1545497424020389248 & DA & 4.7 & 9.71 & 7.32 & 0.19 & 18.68 & 9742 & 8.45 \\
J123414.10+124829.5 & 3931751257626586880 & DA & 3.9 & 2.26 & 11.25 & 0.35 & 17.29 & 8260 & 7.78 \\
J124423.88+292146.8 & 3963112937023073408 & DA & 0.8 & - & 6.21 & -0.23 & 16.85 & 17789 & 7.85 \\
J124836.31+294231.2 & 1465092887460185600 & DA & 2.1 & 2.19 & 15.54 & 0.54 & 17.57 & 6999 & 8.06 \\
J124851.31-022924.7 & 3682578764308891264 & DA & 5.1 & 4.26 & 4.58 & -0.05 & 18.55 & 12282 & 8.12 \\
J125434.65+371000.1 & 1517487399662724608 & DA & 2.3 & 2.64 & 9.55 & -0.26 & 16.18 & 19800 & 8.16 \\
J125434.65+371000.1 & 1517487399662724608 & DA & 1.9 & 2.64 & 9.55 & -0.26 & 16.18 & 19800 & 8.16 \\
J132926.05+254936.5 & 1448232907440917760 & DA & 12.9 & 2.61 & 11.59 & -0.39 & 17.65 & 28122 & 9.48 \\
J134820.79+381017.2 & 1496389390730056576 & DA & 7.7 & 15.52 & 7.18 & -0.43 & 17.81 & 30547 & 9.17 \\
J141906.18+254356.1 & 1258934014870979712 & DA & 1.1 & 2.27 & 13.39 & 0.21 & 17.36 & 9516 & 8.40 \\
J143019.03+281100.5 & 1280674894509973760 & DA & 9 & 7.26 & 14.17 & 0.24 & 17.64 & 9232 & 8.57 \\
J143019.03+281100.5 & 1280674894509973760 & DA & 9.5 & 7.26 & 14.17 & 0.24 & 17.64 & 9232 & 8.57 \\
J150813.24+394504.9 & 1391901769949322624 & DA & 12 & 12.95 & 5.68 & -0.19 & 17.63 & - & - \\
J151325.97+000225.3 & 4418961661806565376 & DA & 5.7 & - & 8.46 & -0.35 & 17.01 & 26520 & 8.79 \\
J151625.07+280320.9 & 1271649969930799872 & DA & 1.2 & 2.3 & 21.24 & 0.40 & 16.48 & 7819 & 8.05 \\
J151625.07+280320.9 & 1271649969930799872 & DA & 1.5 & 2.3 & 21.24 & 0.40 & 16.48 & 7819 & 8.05 \\
J153843.10+084238.2 & 1164767677244452096 & DA & 10.7 & 12.66 & 11.81 & 0.23 & 17.88 & 9265 & 8.49 \\
J154855.07+245113.0 & 1219699145026398848 & DA & 4.2 & 8.41 & 13.61 & -0.27 & 16.80 & 21516 & 9.00 \\
J163604.37+253640.3 & 1300956211117631360 & DA & 1.1 & - & 2.83 & -0.29 & 18.37 & 24451 & 8.08 \\
J170751.98+353239.6 & 1338455643596995072 & DA & 1.8 & 2.64 & 13.29 & -0.31 & 16.49 & 23440 & 8.91 \\
J215843.43+052741.3 & 2697238362376404224 & DA & 2.1 & - & 11.50 & 0.12 & 17.60 & 10371 & 8.51 \\
J224741.46+145638.7 & 2732459327587247360 & DQ & - & 515.09 & 9.77 & -0.17 & 17.52 & 16452 & 8.76 \\
J235430.19+343745.6 & 2878389676217300736 & DA & 4.3 & - & 9.95 & 0.10 & 17.73 & 10926 & 8.47 \\
J235430.24+343745.3 & 2878389676217300736 & DA & 4.2 & - & 9.95 & 0.10 & 17.73 & 10926 & 8.47 \\

\end{longtable}

\newpage

\section{Atlas of Magnetic White Dwarf Spectra}
\label{appendix:spectra}

In this appendix, we present the full sample of MWDs discovered in this work. For DA MWDs, the rest-frame positions of the Balmer series (H$\alpha$, H$\beta$, and H$\gamma$) are indicated by red dashed vertical lines. For helium-rich (DB) MWDs, blue dotted lines indicate the rest-frame positions of prominent He~I transitions. In cases where the spectrum is dominated by carbon features (DQ MWDs), no reference lines are plotted. The mean surface magnetic field ($B_{\text{mean}}$) is provided for confirmed targets alongside the LAMOST designation.

\begin{figure*}
    \centering
\includegraphics[width=8.5cm]{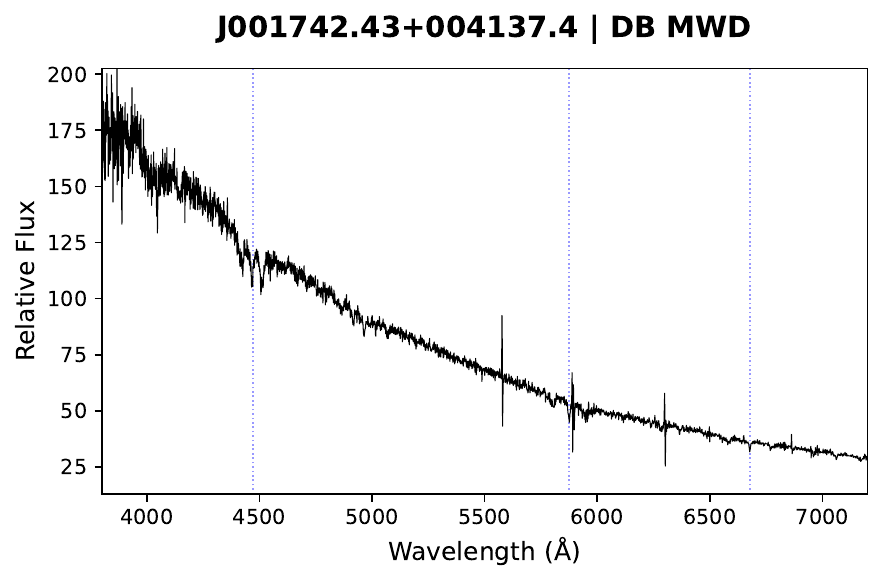}
\includegraphics[width=8.5cm]{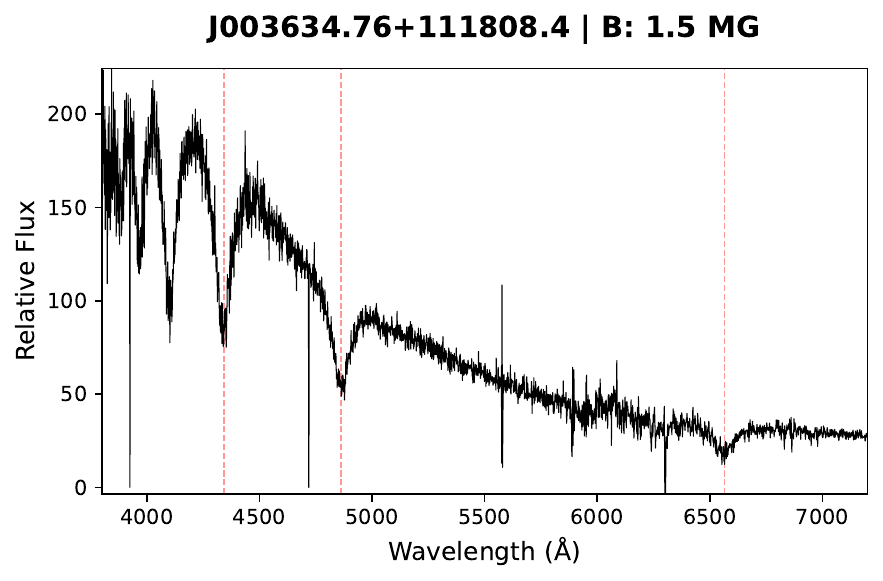}
\includegraphics[width=8.5cm]{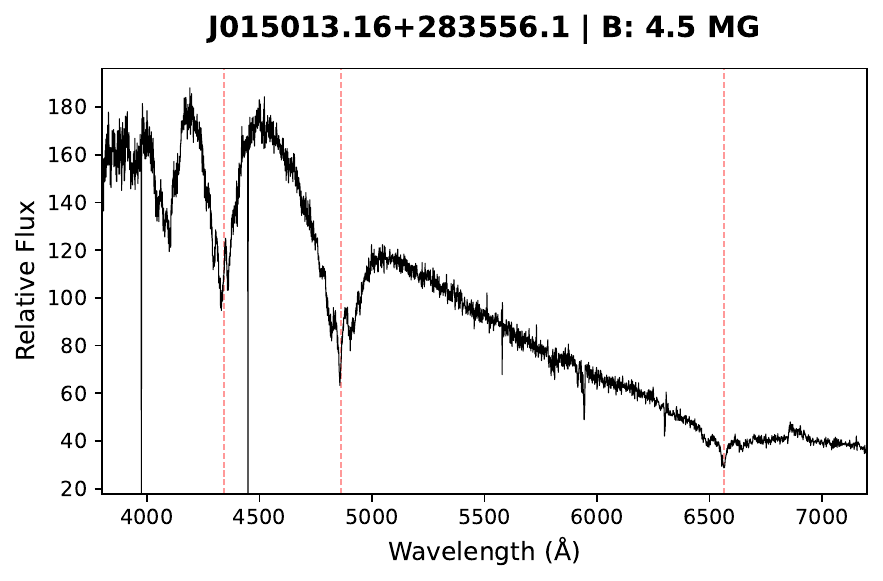}
\includegraphics[width=8.5cm]{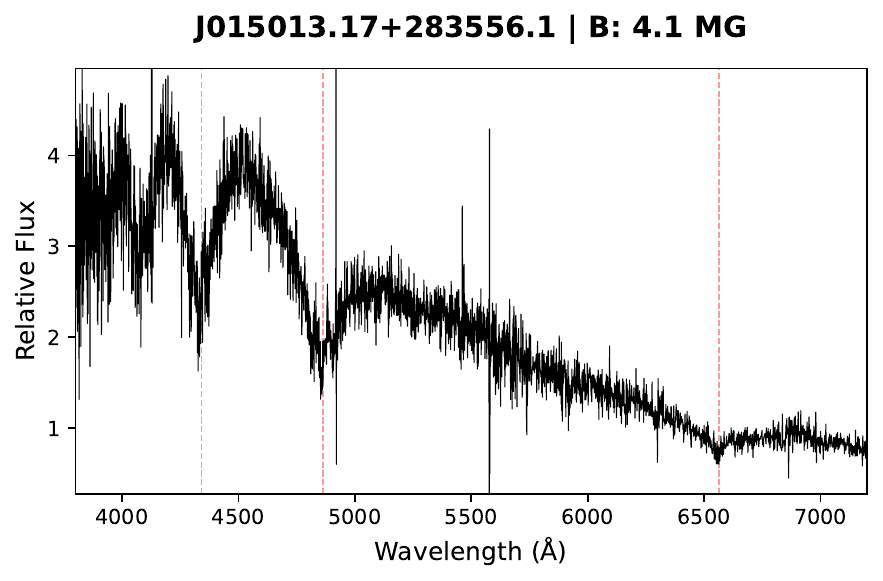}
\includegraphics[width=8.5cm]{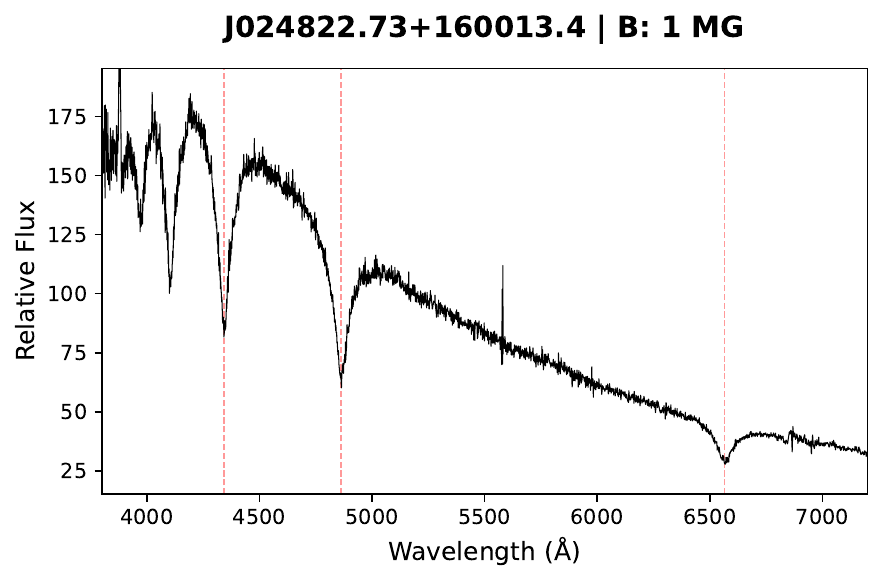}
\includegraphics[width=8.5cm]{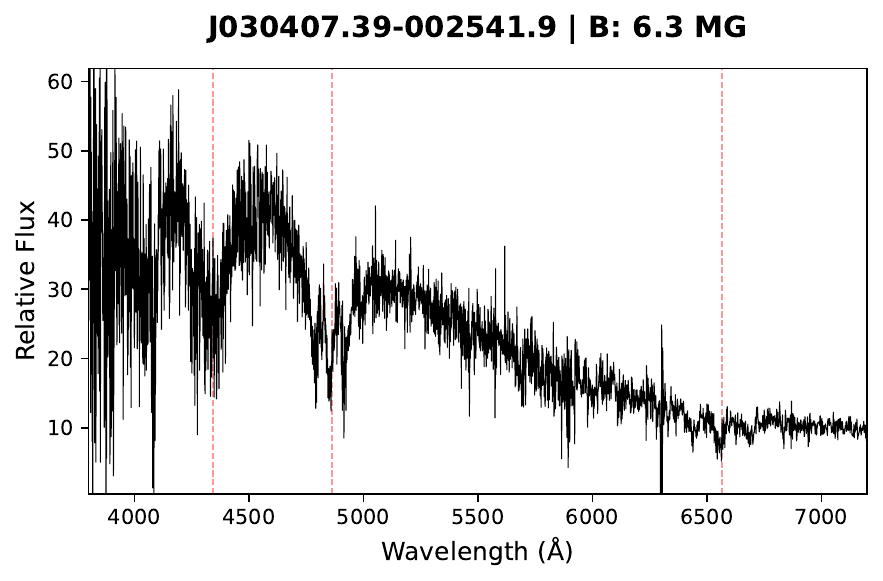}
\end{figure*}

\begin{figure*}
    \centering
\includegraphics[width=8.5cm]{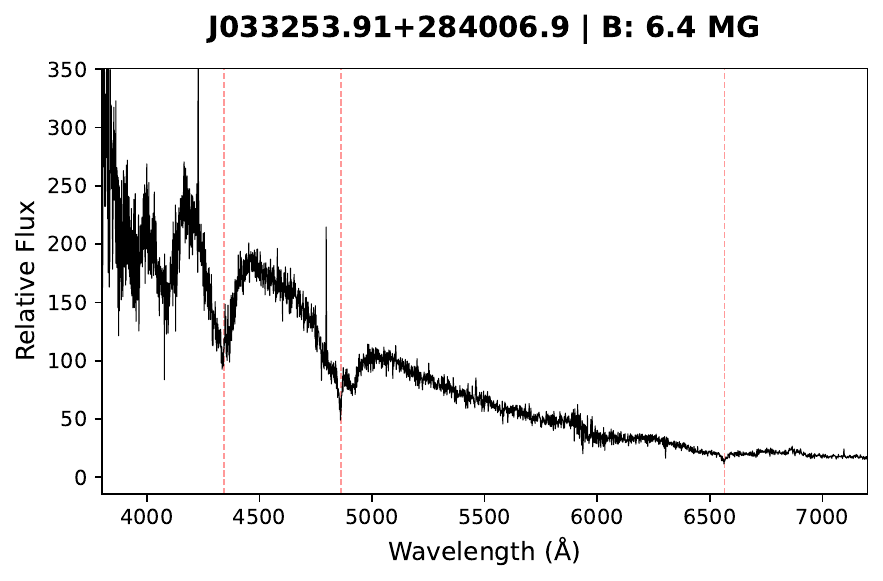}
\includegraphics[width=8.5cm]{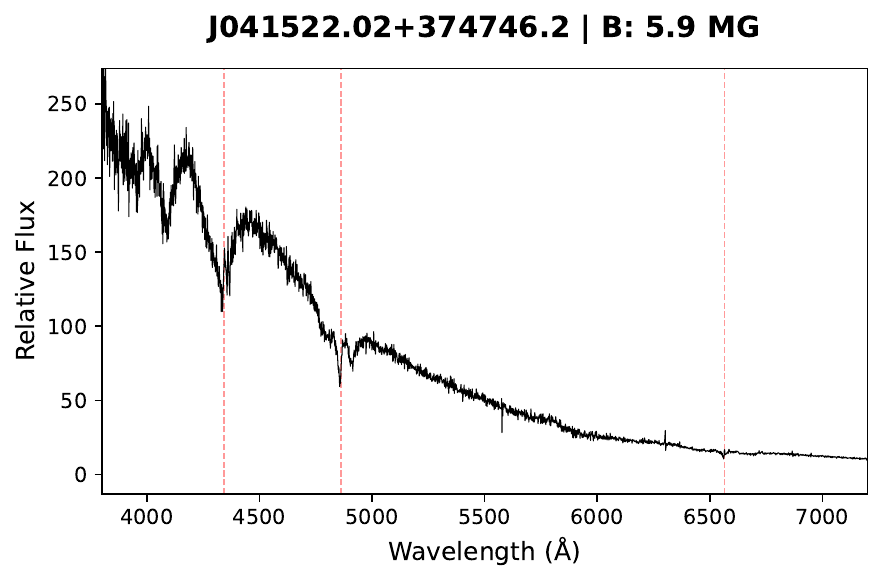}
\includegraphics[width=8.5cm]{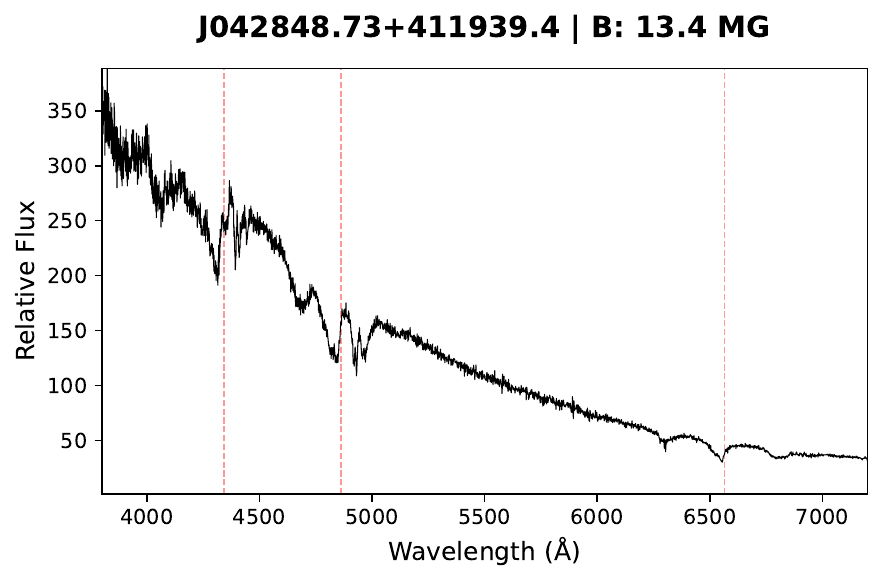}
\includegraphics[width=8.5cm]{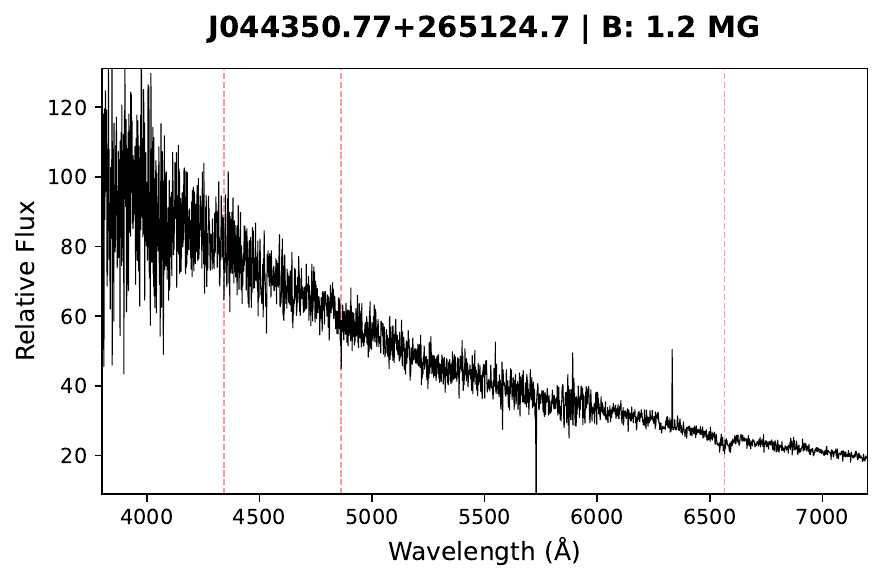}
\includegraphics[width=8.5cm]{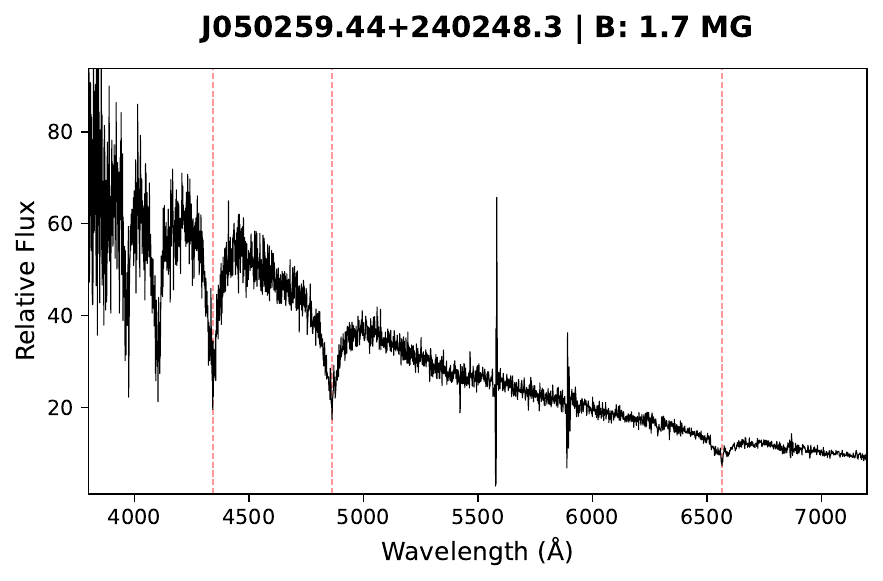}
\includegraphics[width=8.5cm]{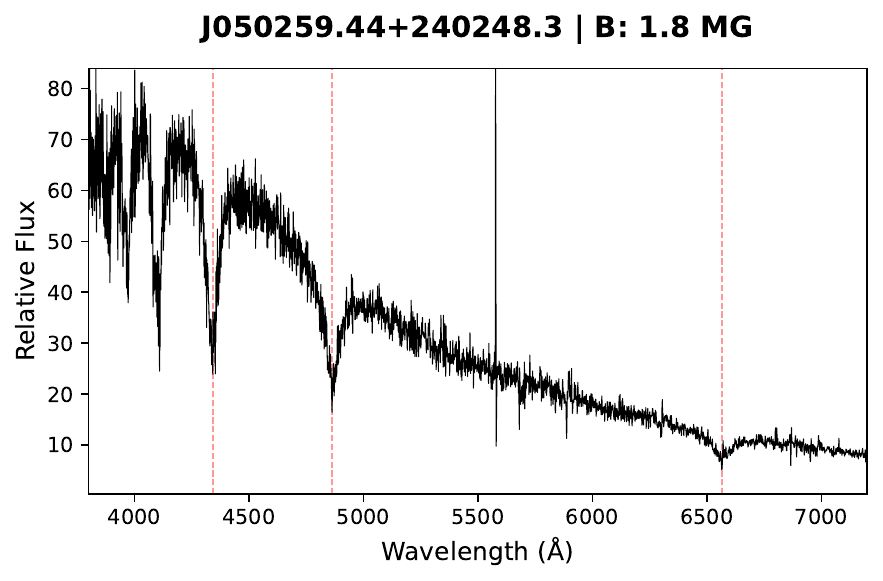}
\includegraphics[width=8.5cm]{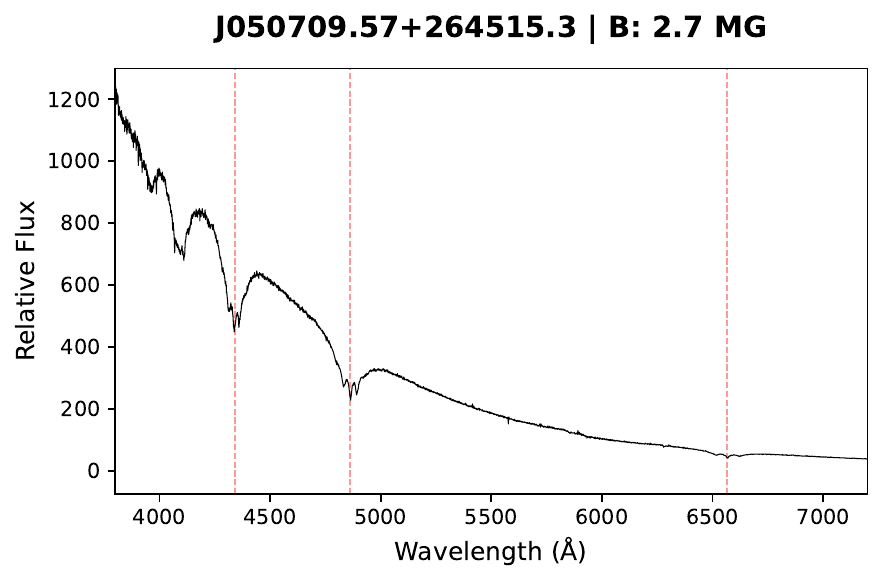}
\includegraphics[width=8.5cm]{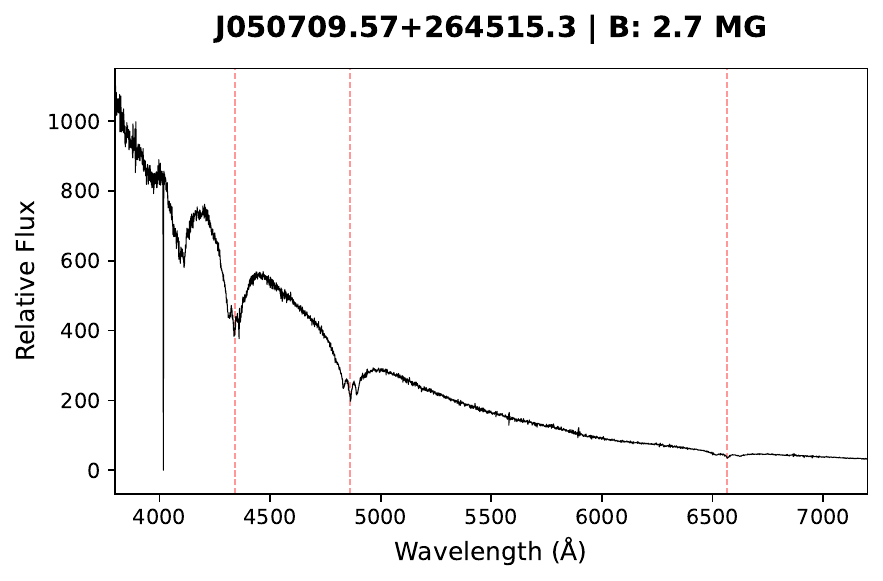}
\end{figure*}

\begin{figure*}
    \centering
\includegraphics[width=8.5cm]{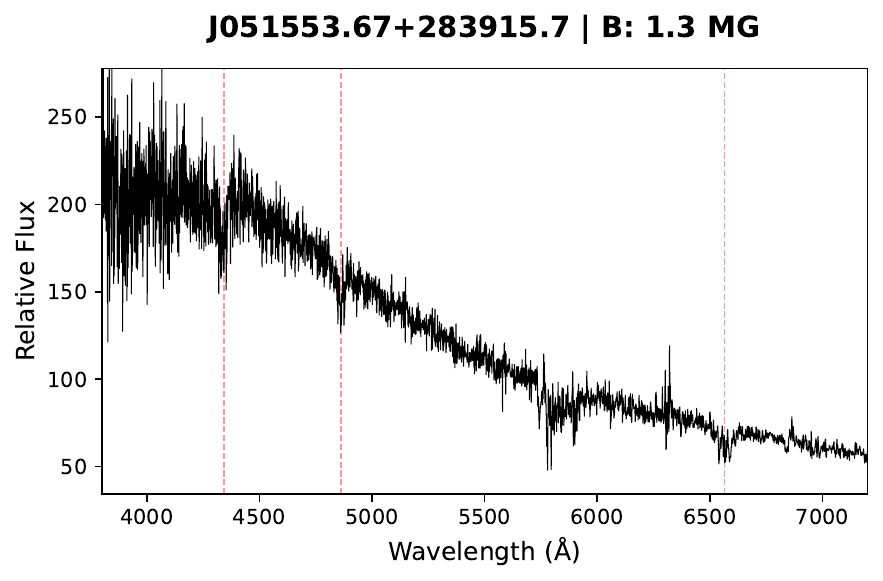}
\includegraphics[width=8.5cm]{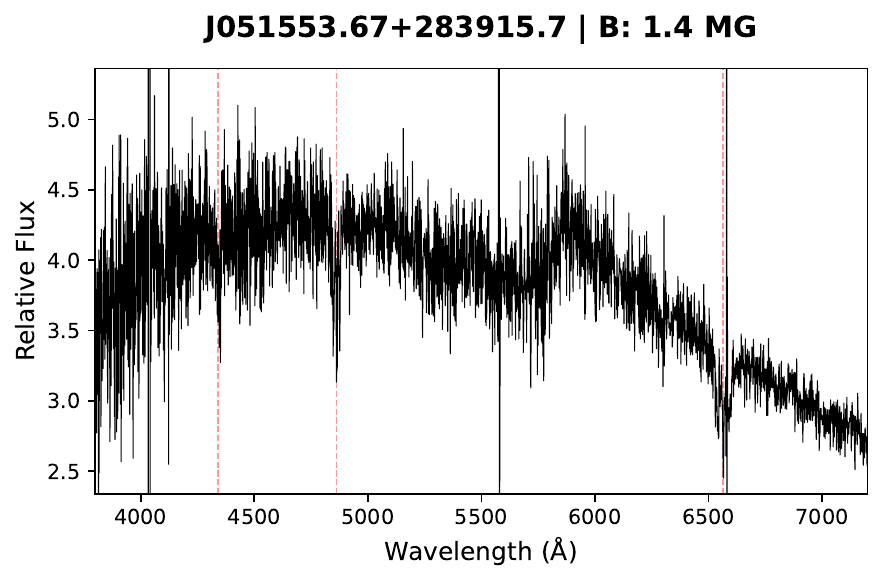}
\includegraphics[width=8.5cm]{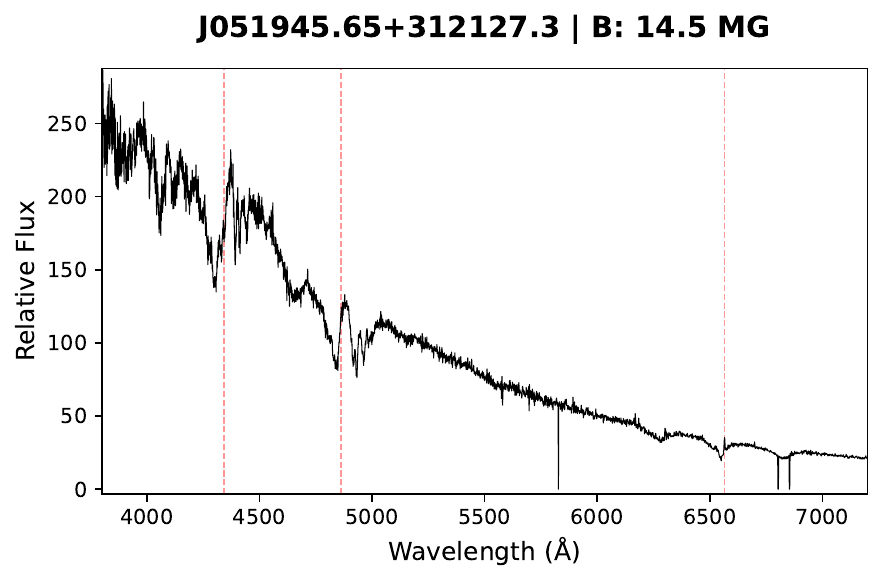}
\includegraphics[width=8.5cm]{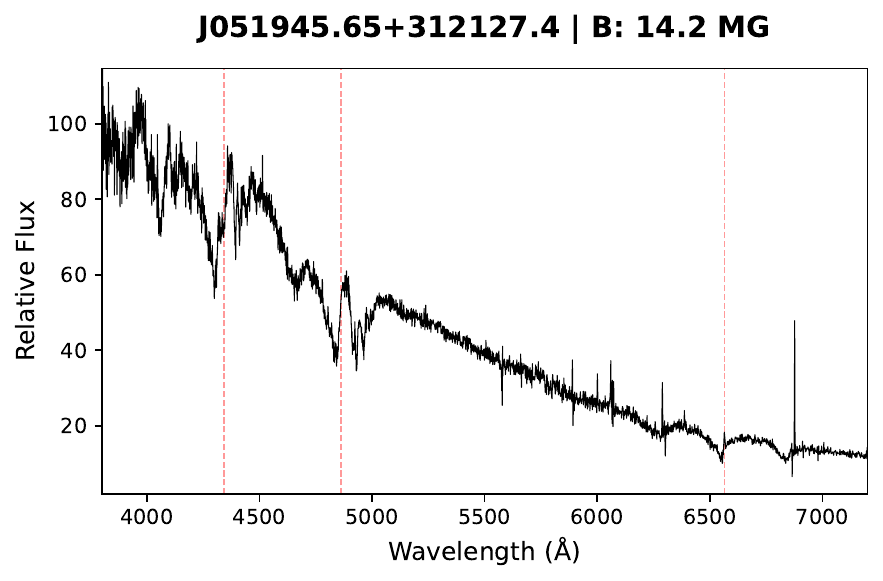}
\includegraphics[width=8.5cm]{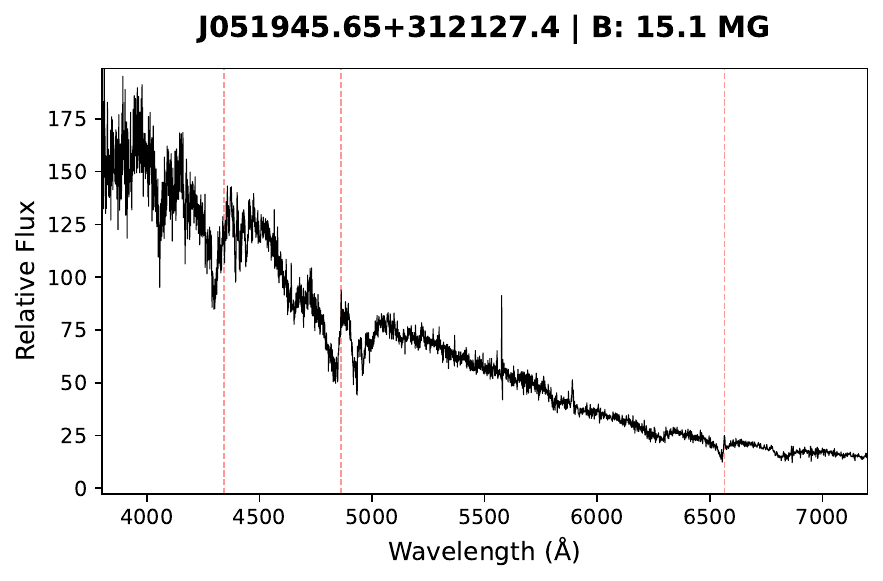}
\includegraphics[width=8.5cm]{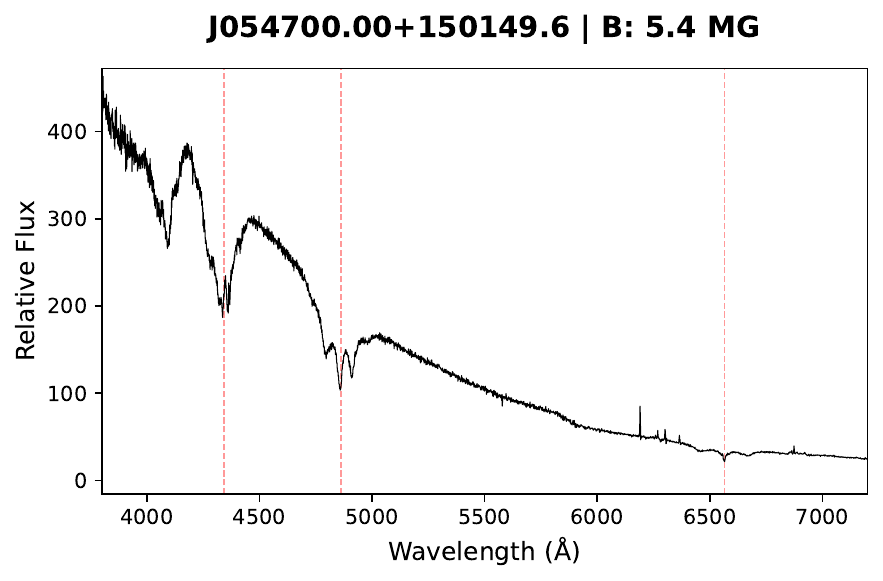}
\includegraphics[width=8.5cm]{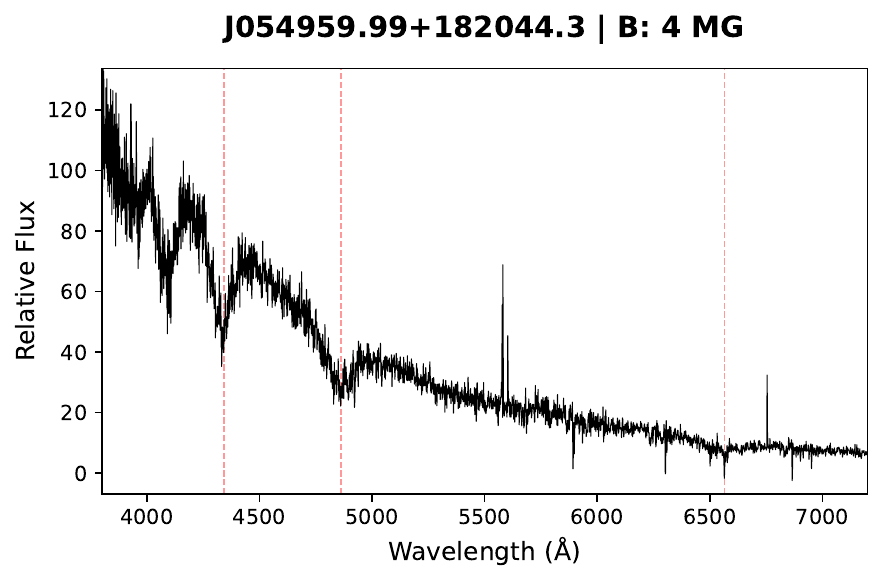}
\includegraphics[width=8.5cm]{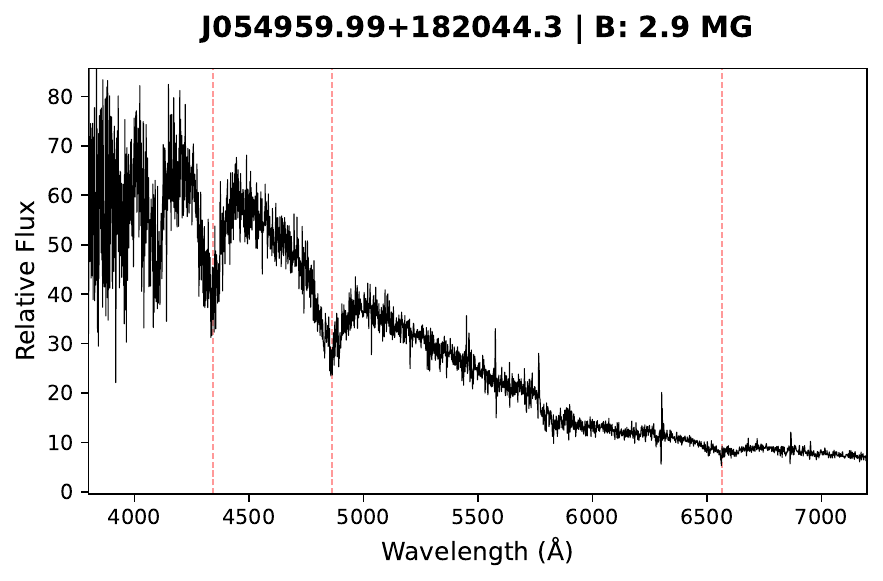}
\end{figure*}

\begin{figure*}
    \centering
\includegraphics[width=8.5cm]{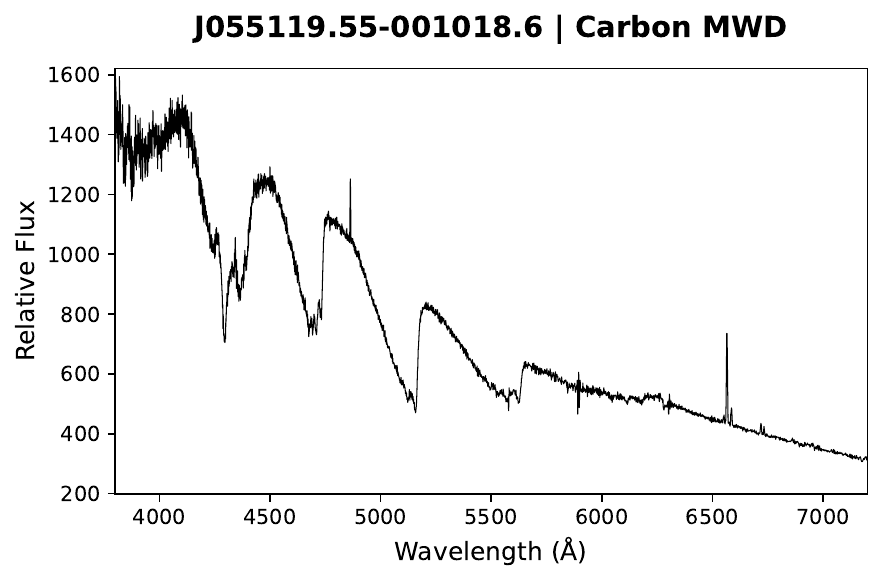}
\includegraphics[width=8.5cm]{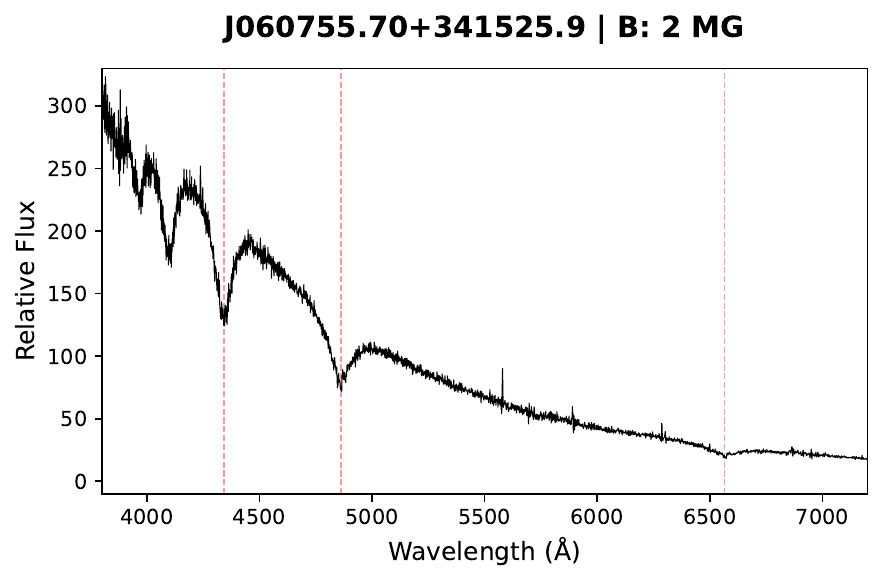}
\includegraphics[width=8.5cm]{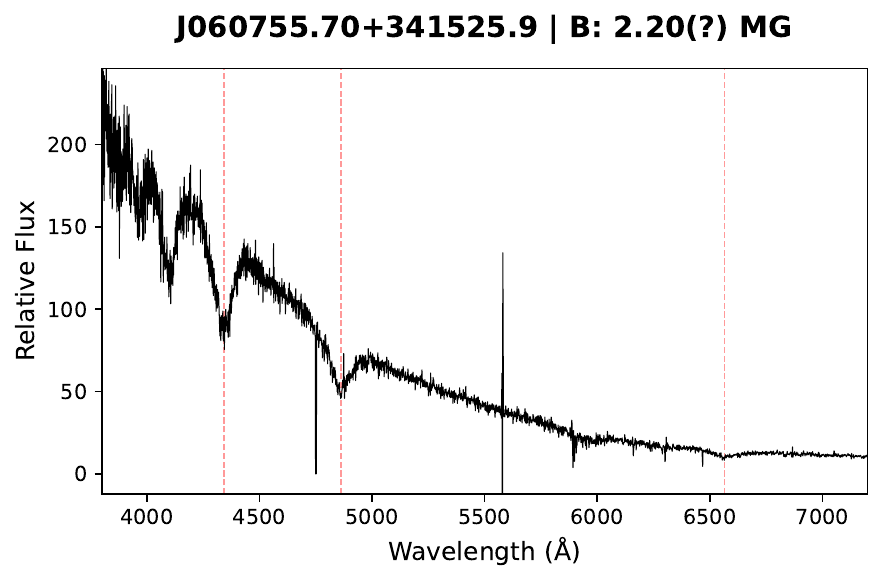}
\includegraphics[width=8.5cm]{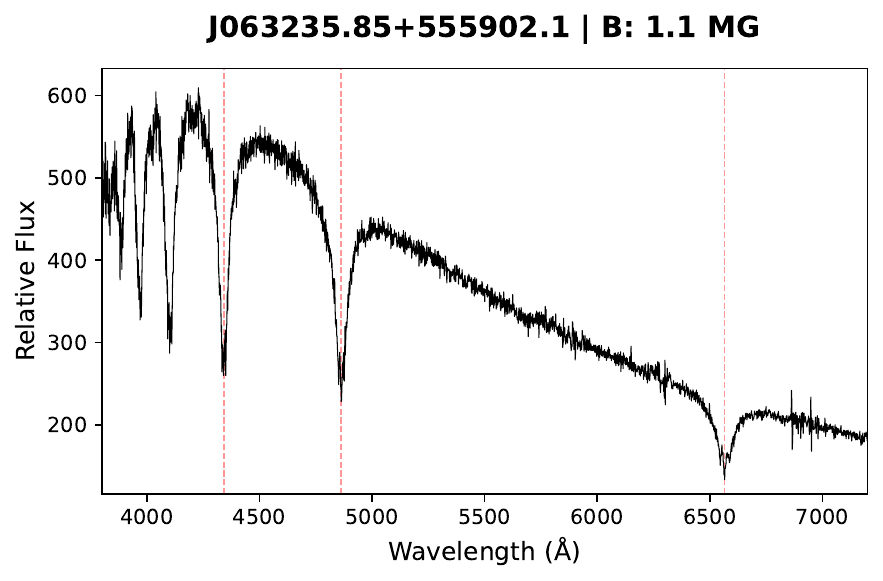}
\includegraphics[width=8.5cm]{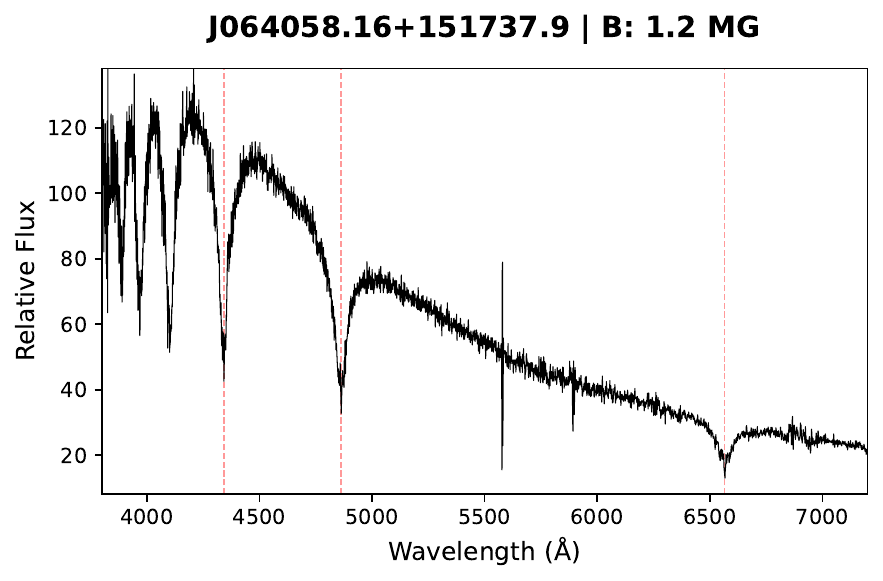}
\includegraphics[width=8.5cm]{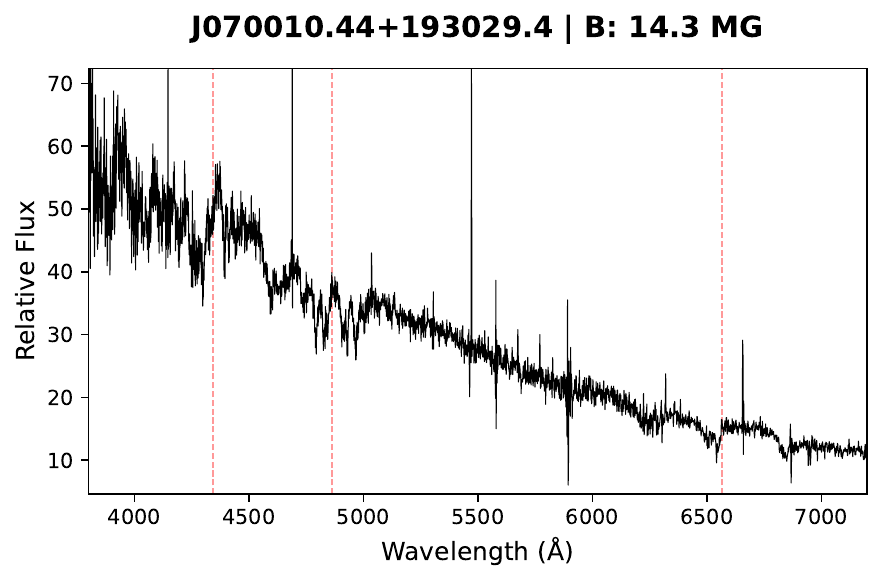}
\includegraphics[width=8.5cm]{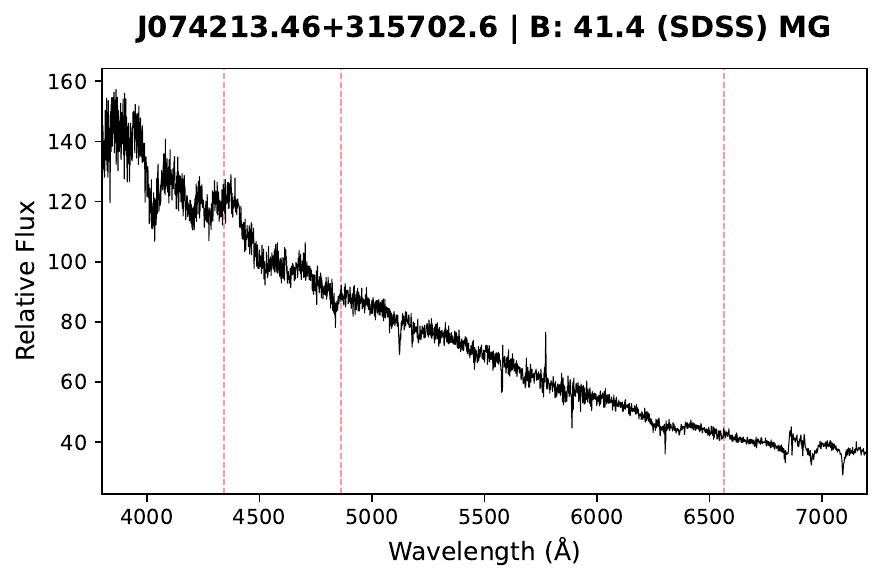}
\includegraphics[width=8.5cm]{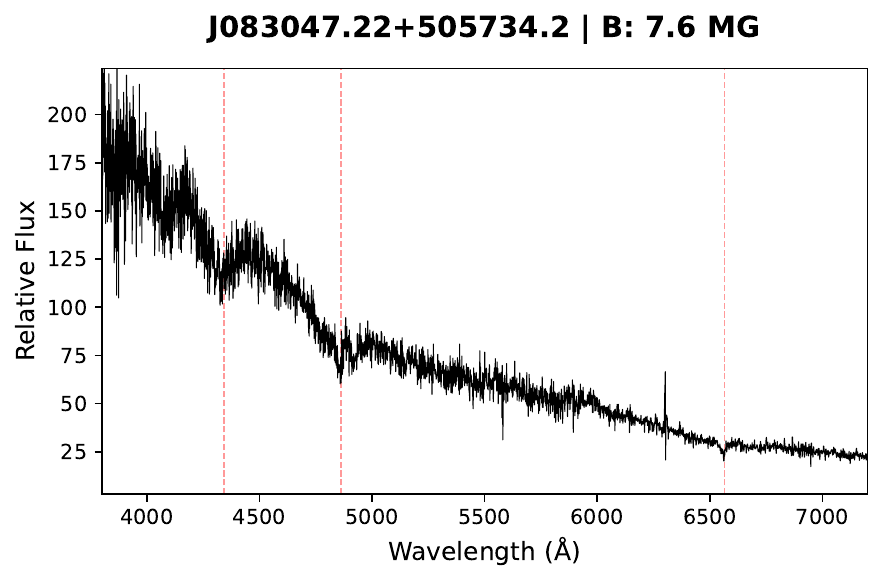}
\end{figure*}

\begin{figure*}
    \centering
\includegraphics[width=8.5cm]{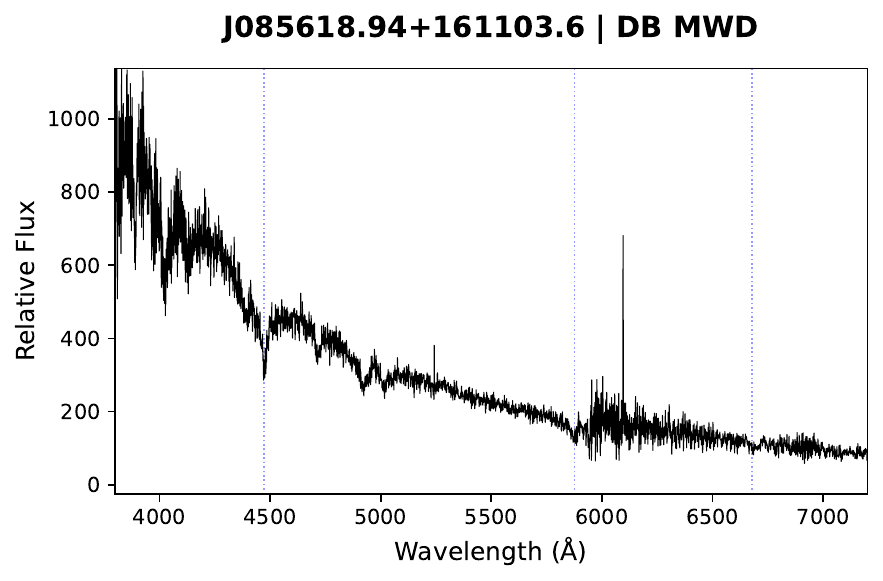}
\includegraphics[width=8.5cm]{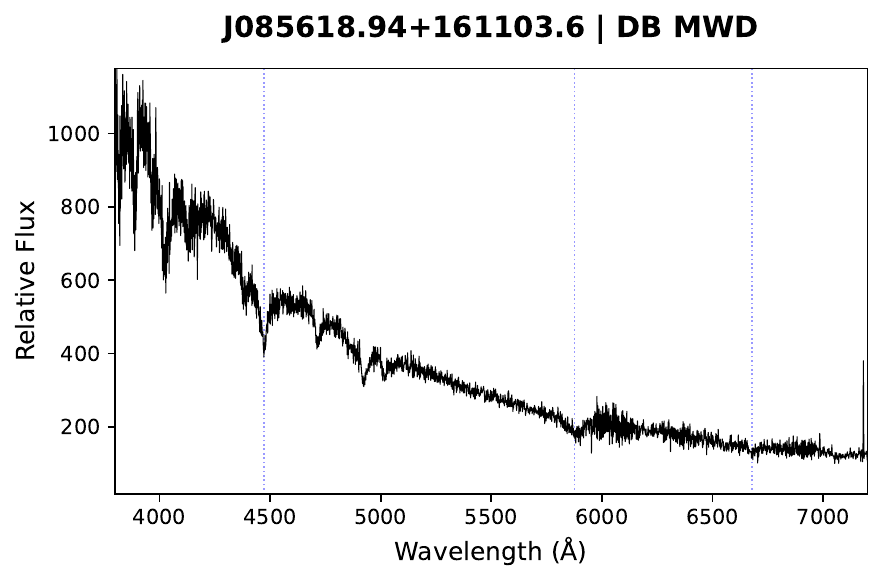}
\includegraphics[width=8.5cm]{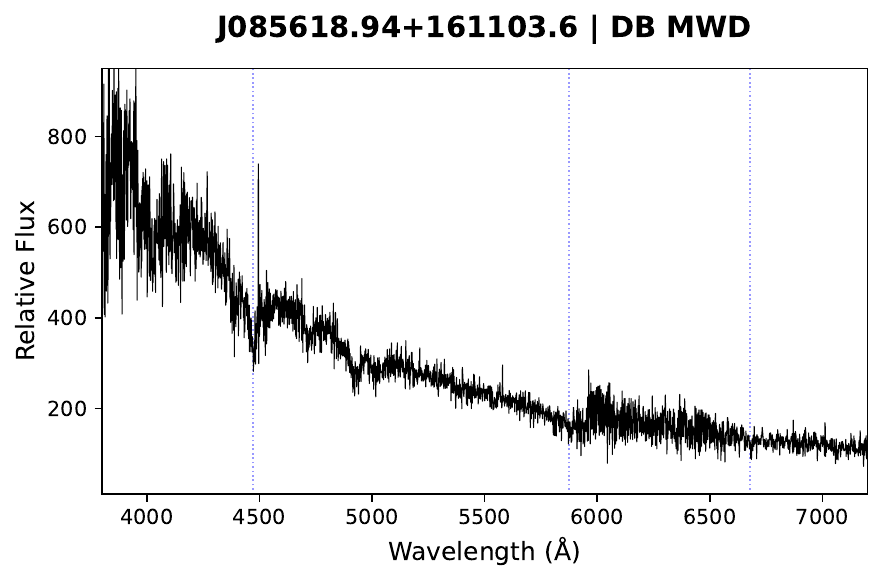}
\includegraphics[width=8.5cm]{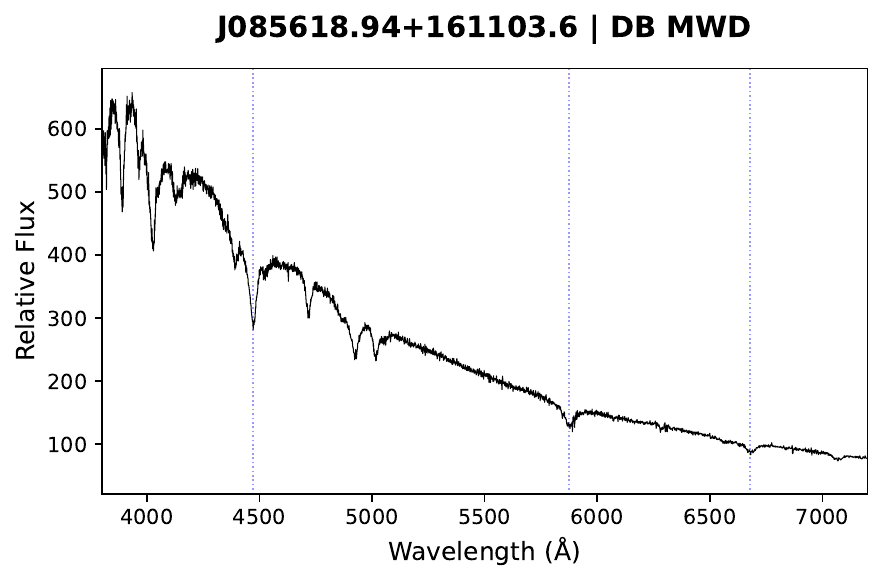}
\includegraphics[width=8.5cm]{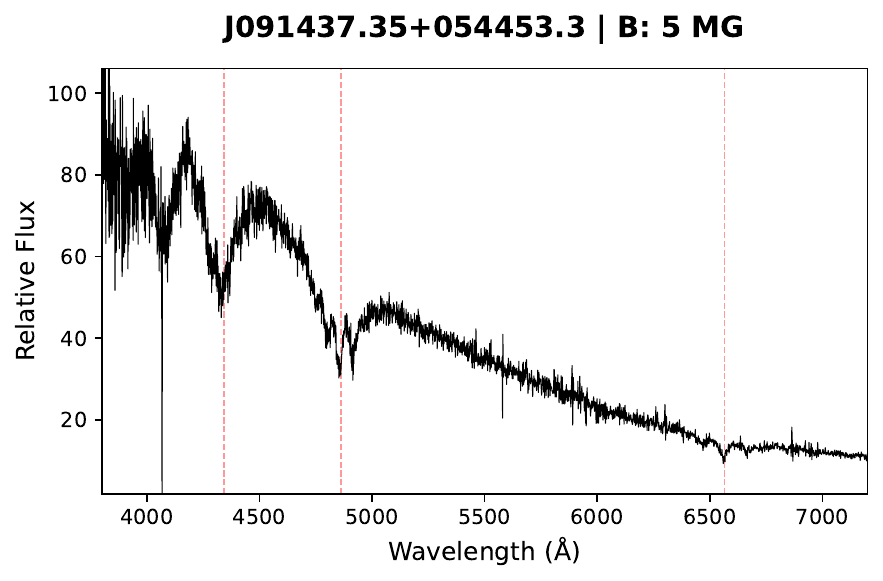}
\includegraphics[width=8.5cm]{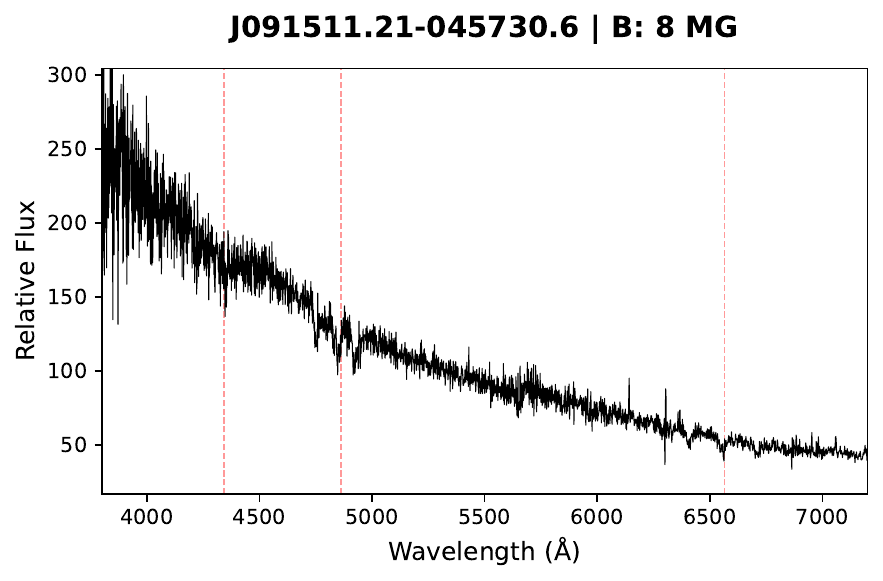}
\includegraphics[width=8.5cm]{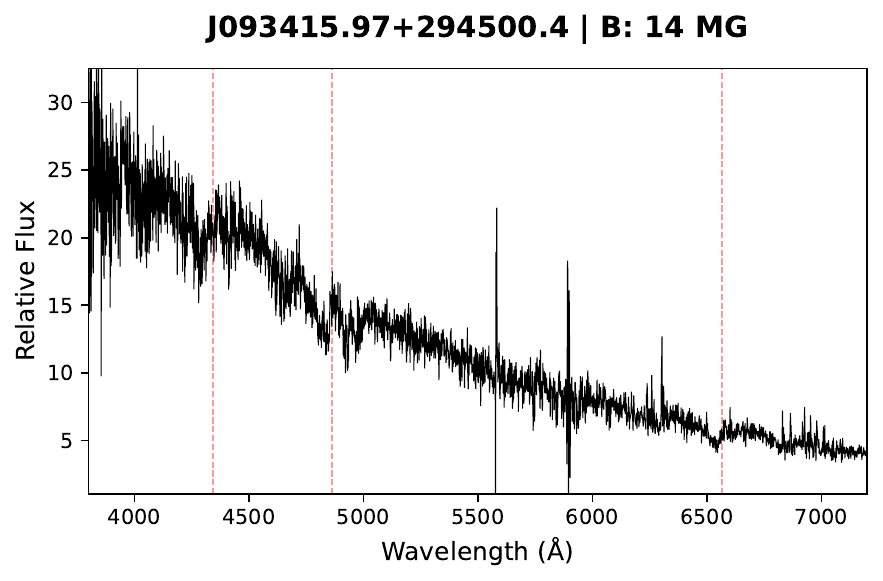}
\includegraphics[width=8.5cm]{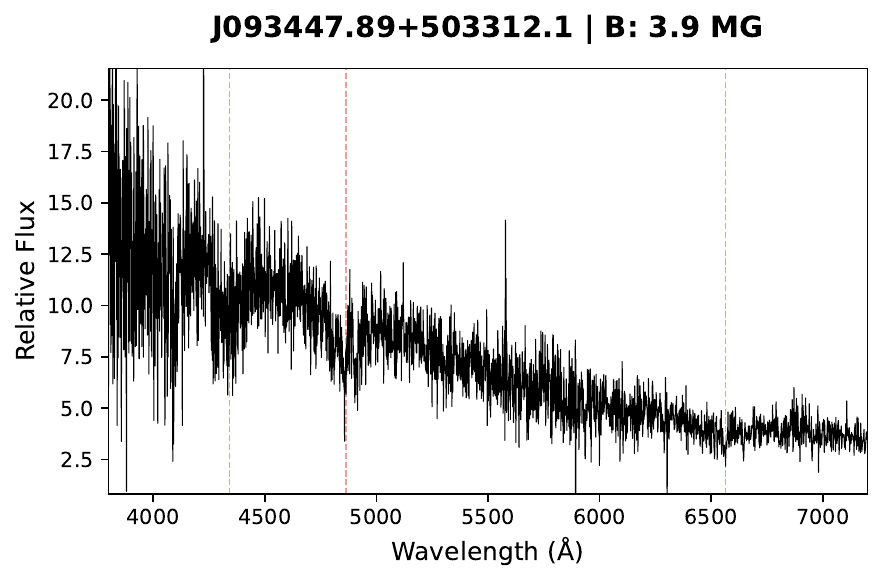}
\end{figure*}

\begin{figure*}
    \centering
\includegraphics[width=8.5cm]{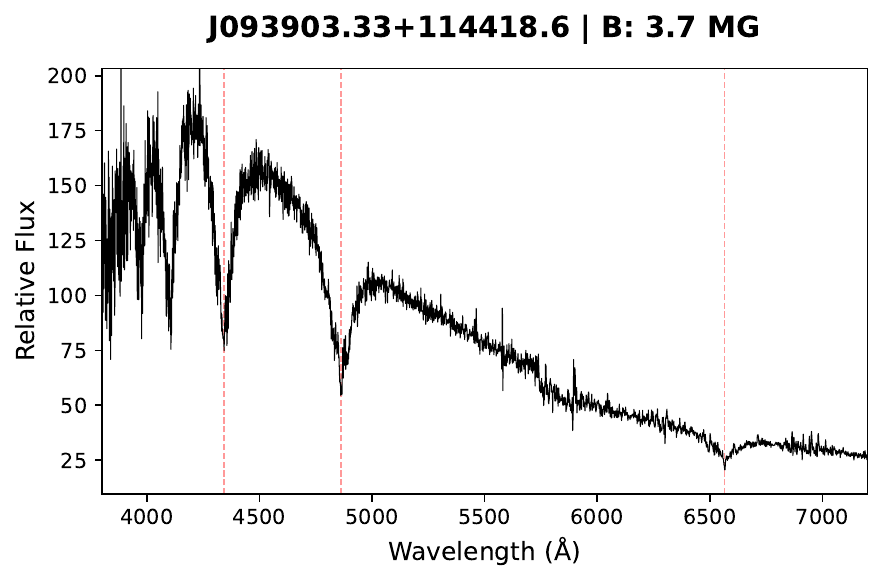}
\includegraphics[width=8.5cm]{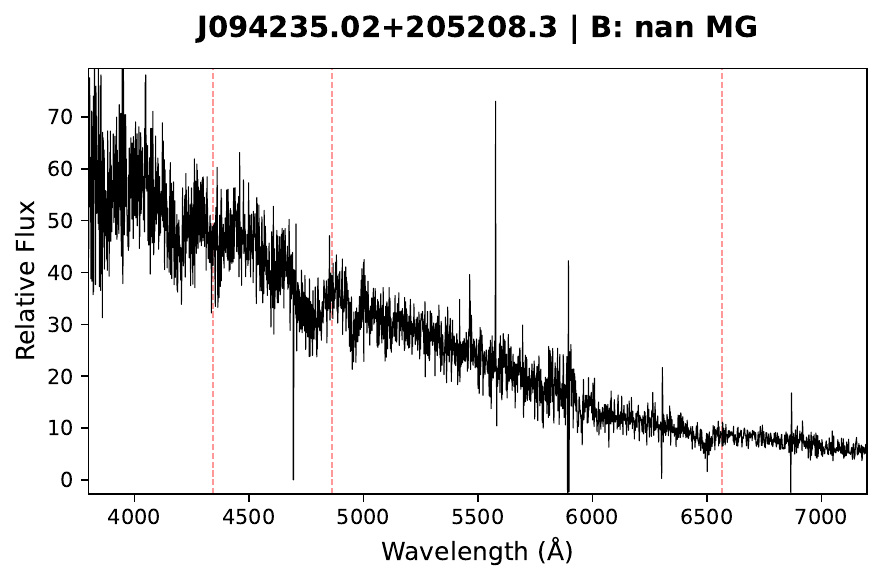}
\includegraphics[width=8.5cm]{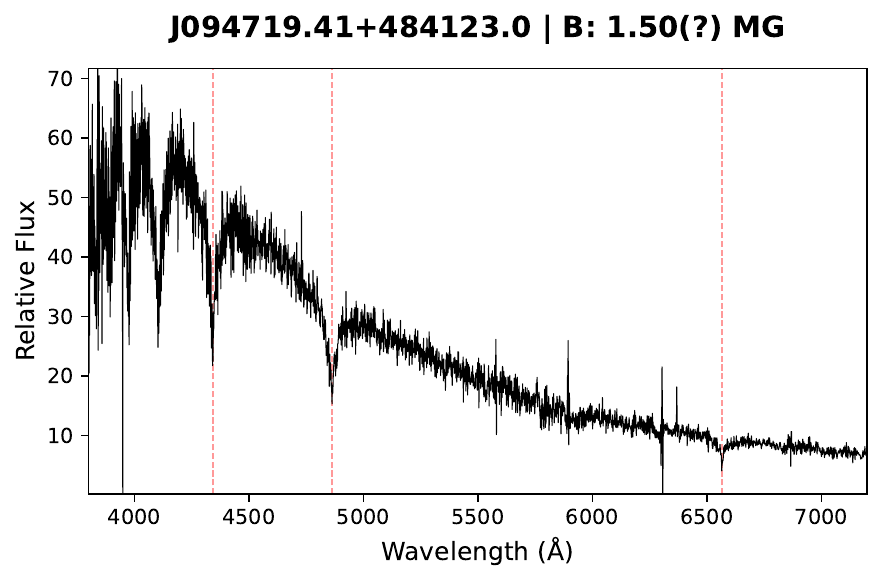}
\includegraphics[width=8.5cm]{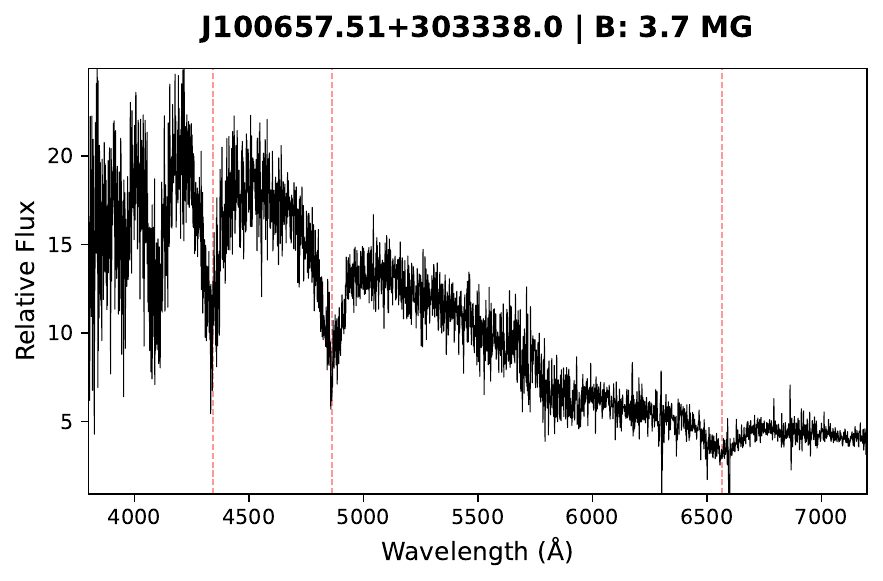}
\includegraphics[width=8.5cm]{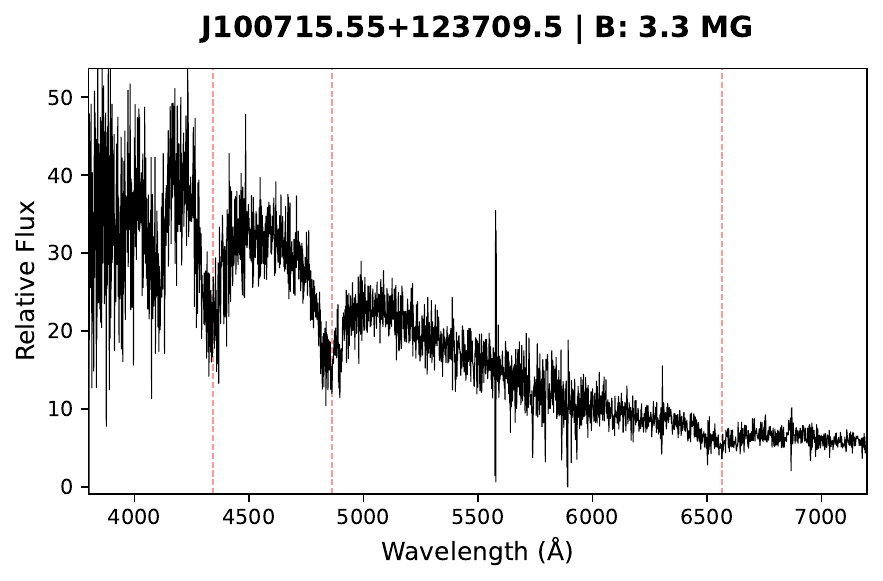}
\includegraphics[width=8.5cm]{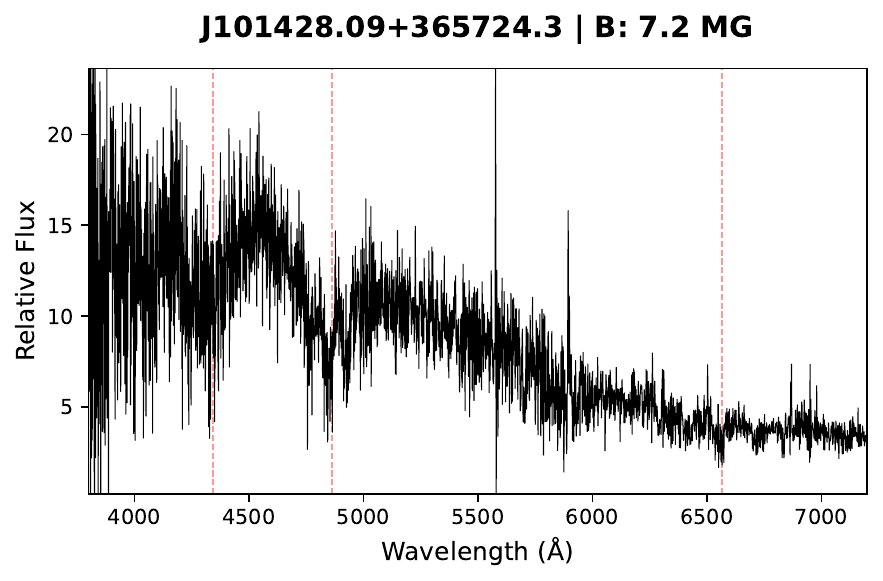}
\includegraphics[width=8.5cm]{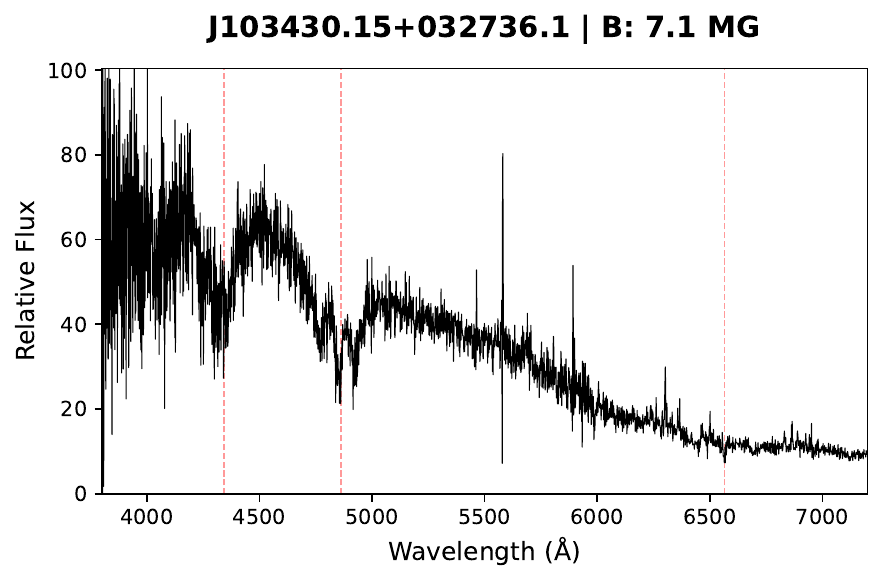}
\includegraphics[width=8.5cm]{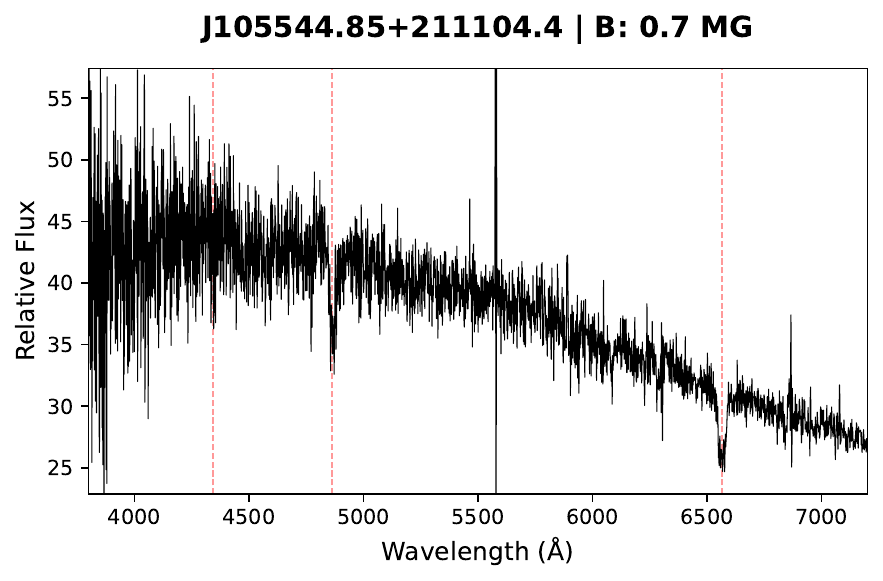}
\end{figure*}

\begin{figure*}
    \centering
\includegraphics[width=8.5cm]{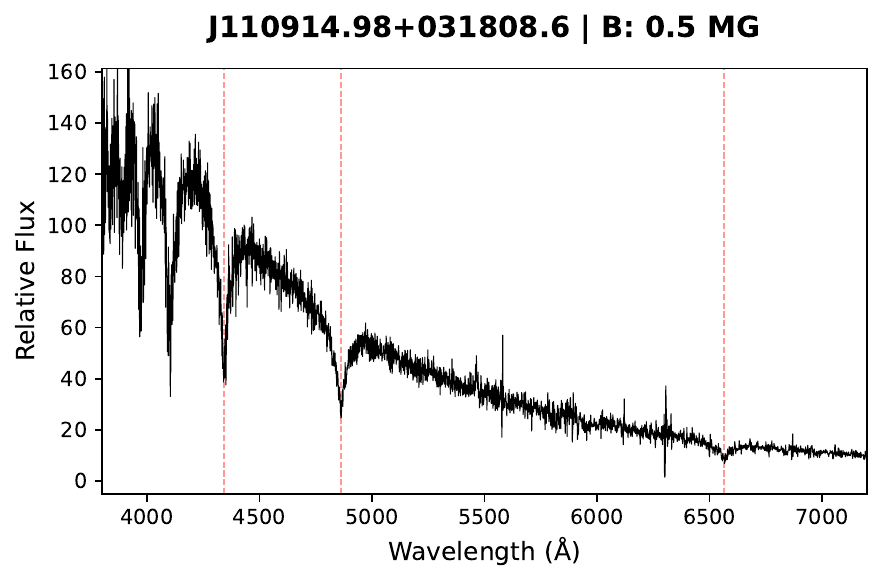}
\includegraphics[width=8.5cm]{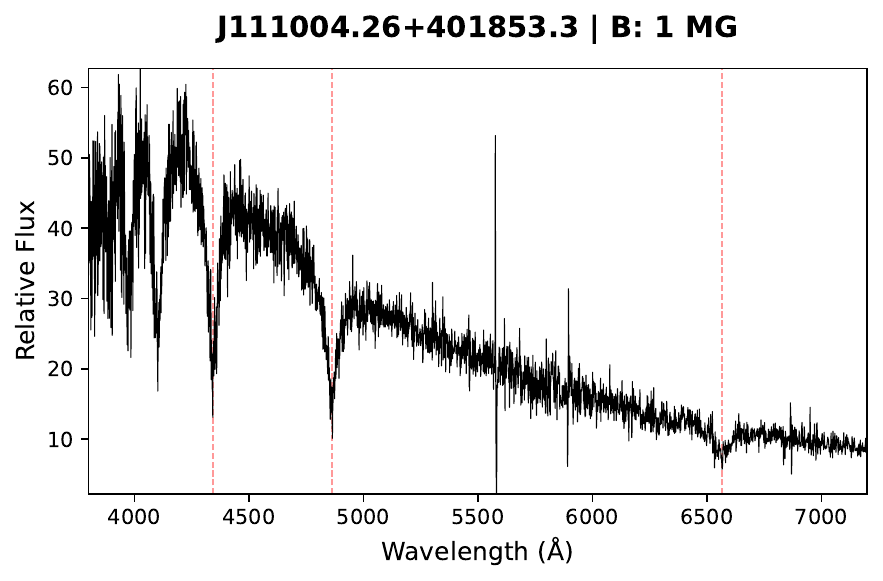}
\includegraphics[width=8.5cm]{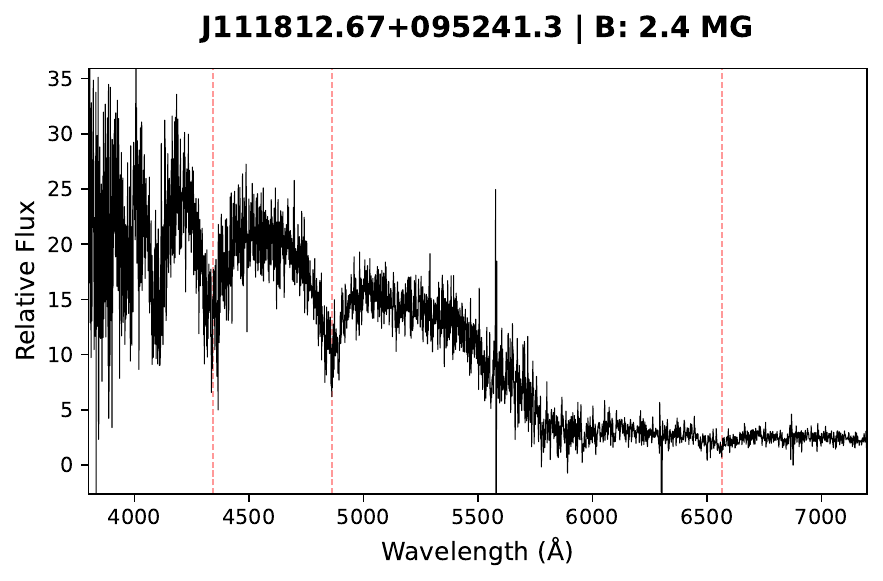}
\includegraphics[width=8.5cm]{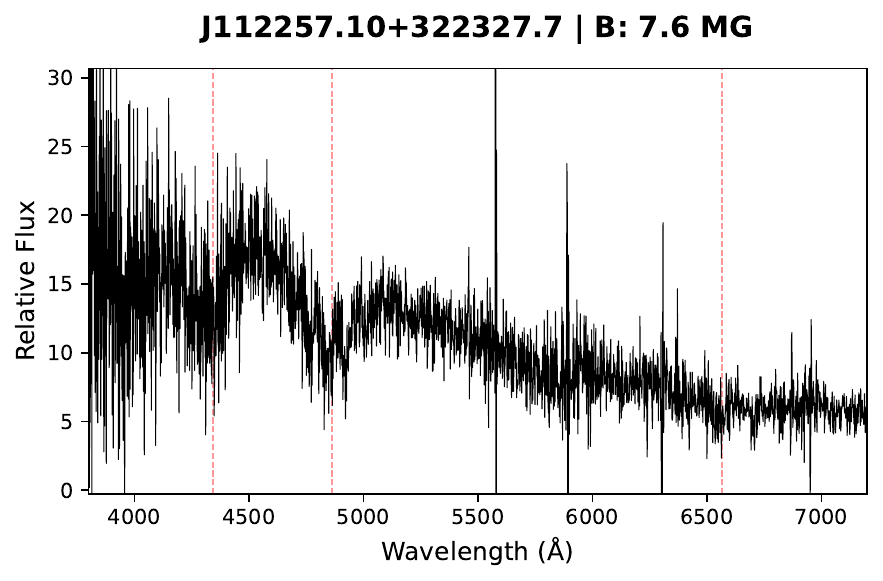}
\includegraphics[width=8.5cm]{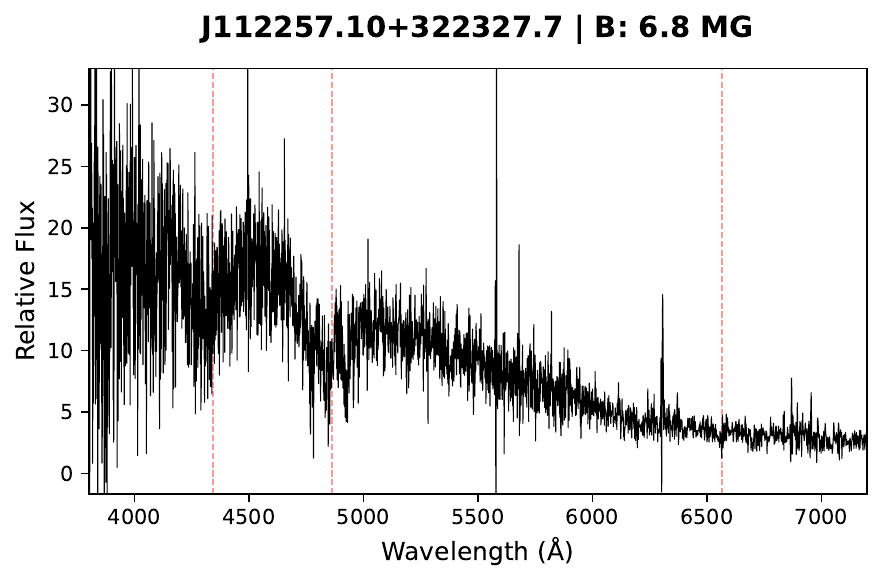}
\includegraphics[width=8.5cm]{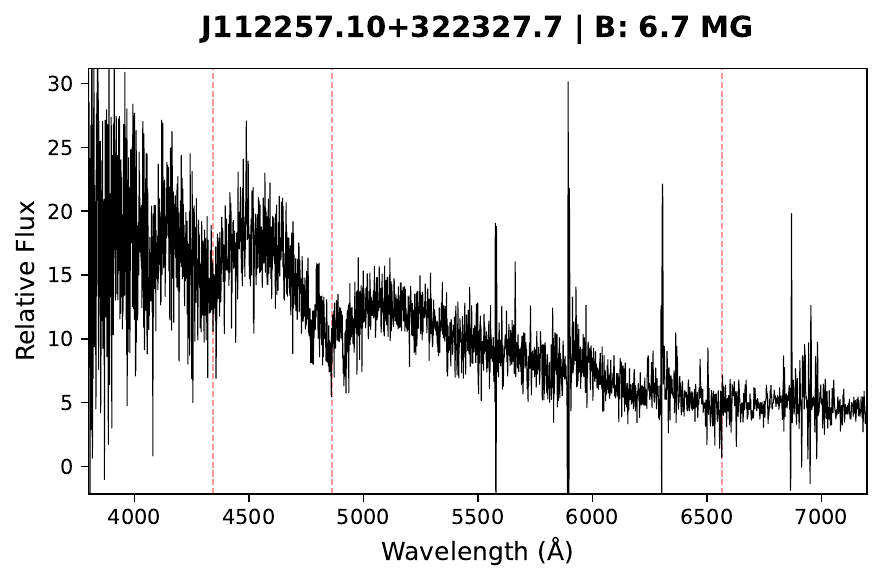}
\includegraphics[width=8.5cm]{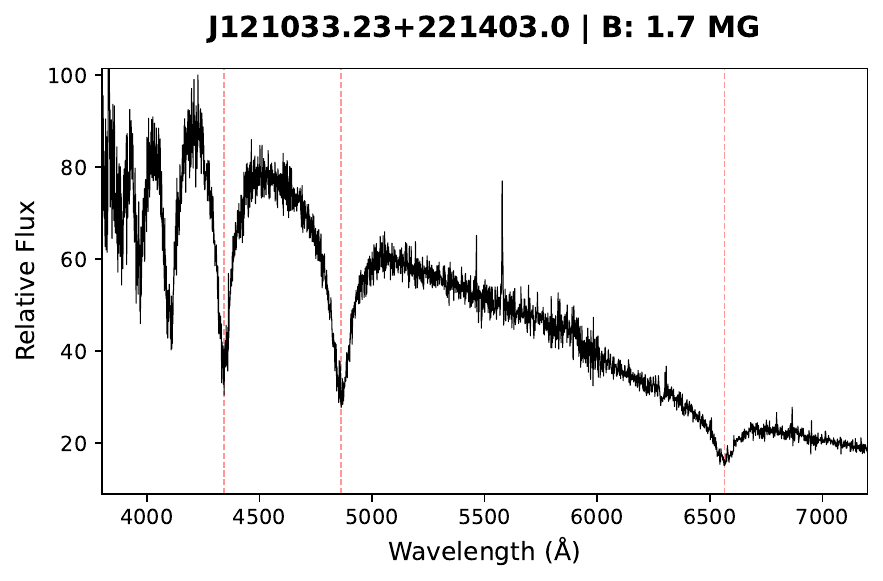}
\includegraphics[width=8.5cm]{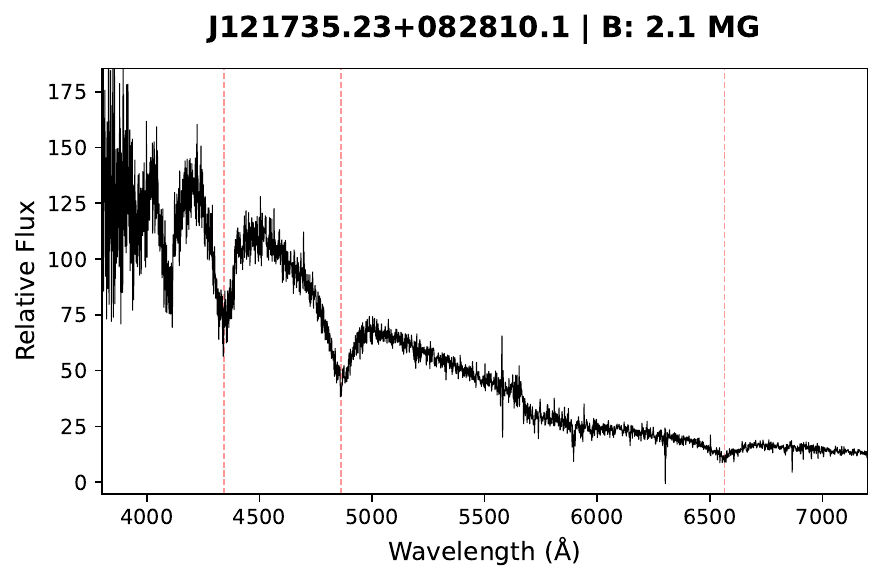}
\end{figure*}

\begin{figure*}
    \centering
\includegraphics[width=8.5cm]{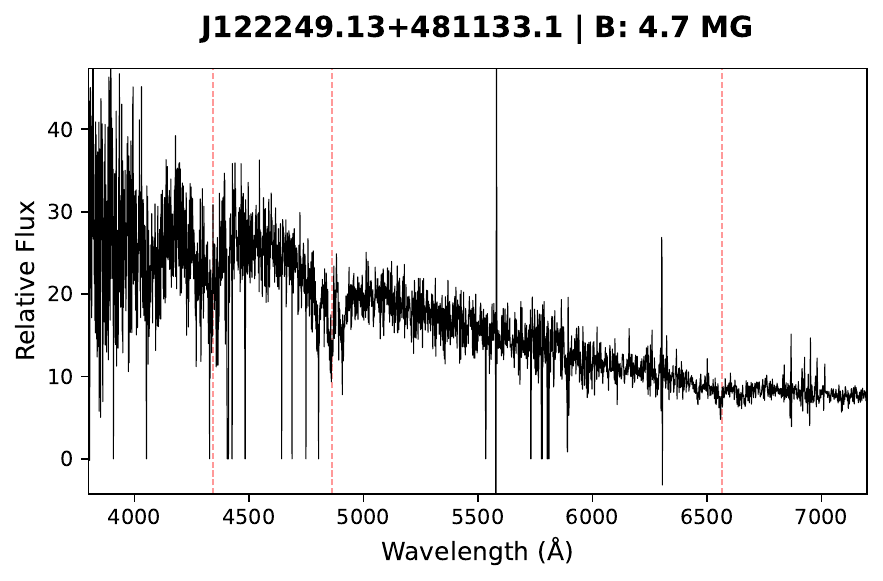}
\includegraphics[width=8.5cm]{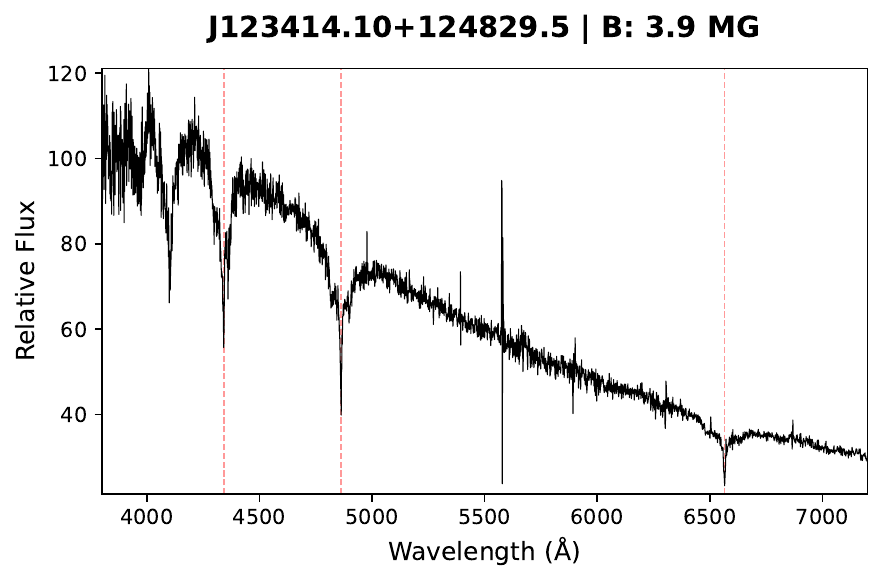}
\includegraphics[width=8.5cm]{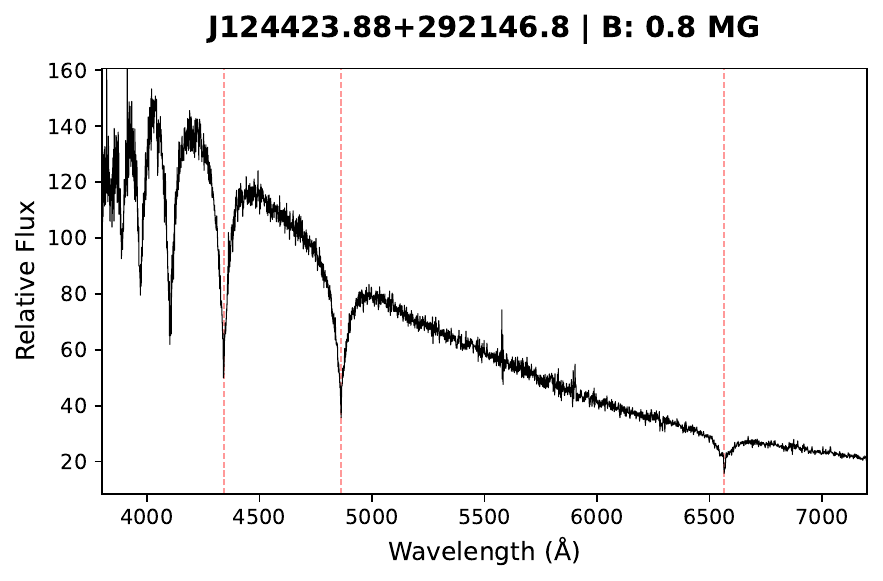}
\includegraphics[width=8.5cm]{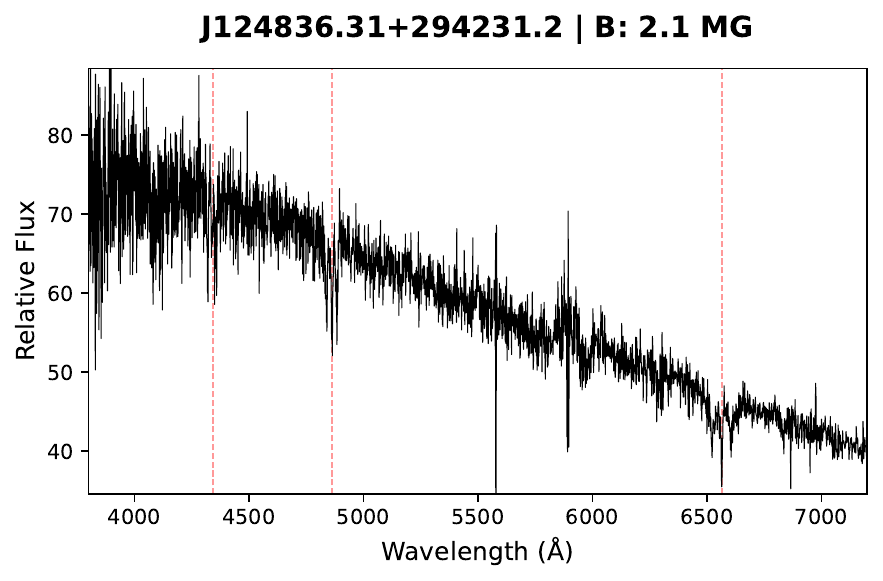}
\includegraphics[width=8.5cm]{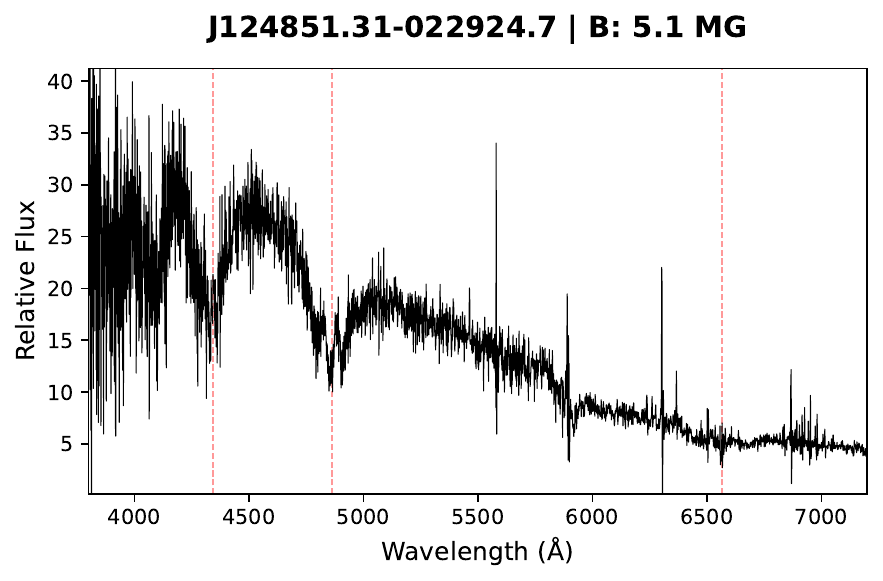}
\includegraphics[width=8.5cm]{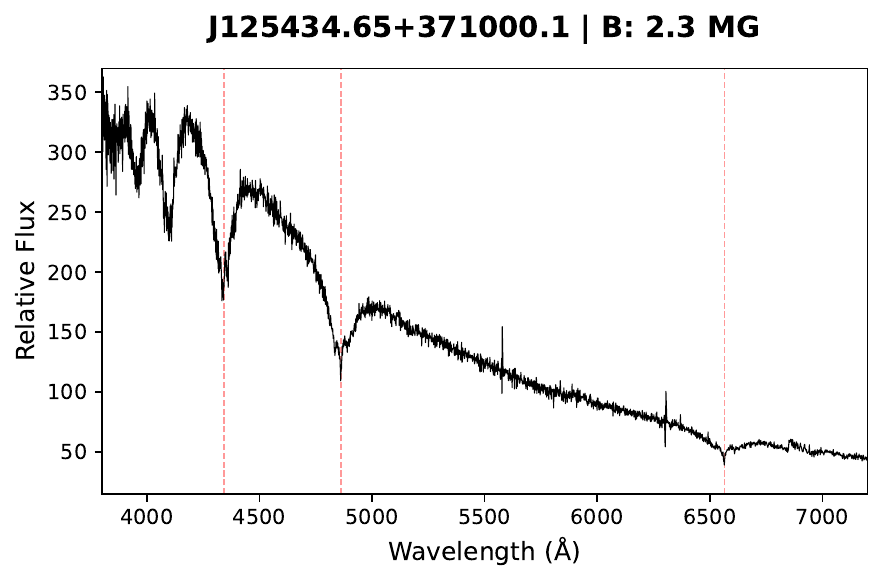}
\includegraphics[width=8.5cm]{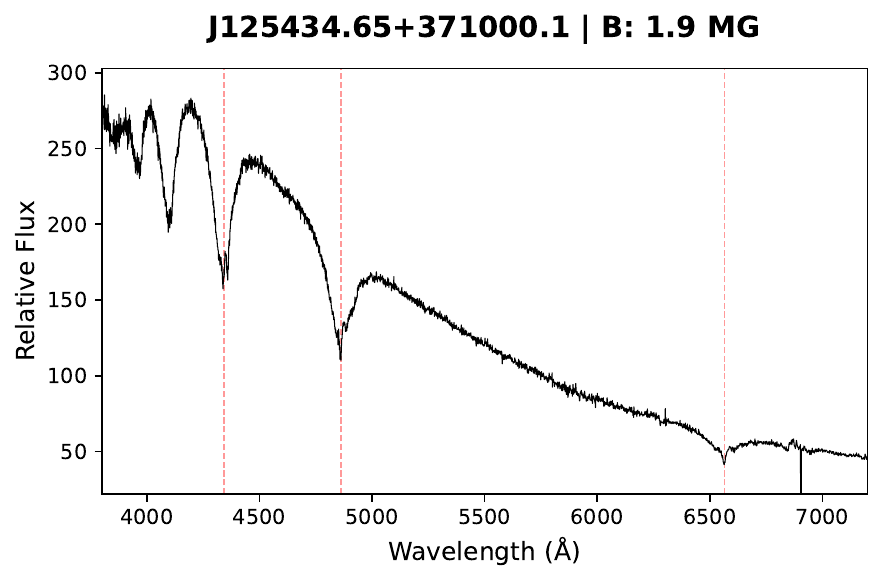}
\includegraphics[width=8.5cm]{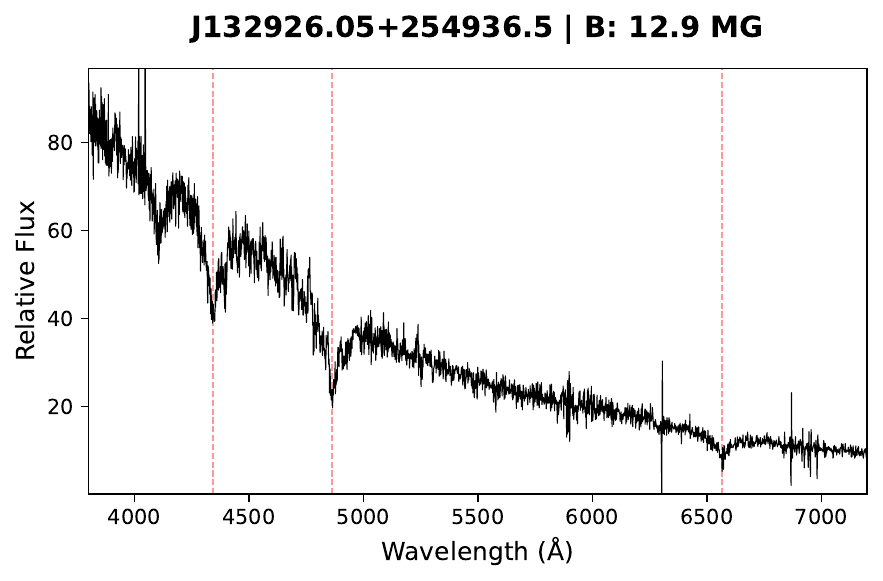}
\end{figure*}

\begin{figure*}
    \centering
\includegraphics[width=8.5cm]{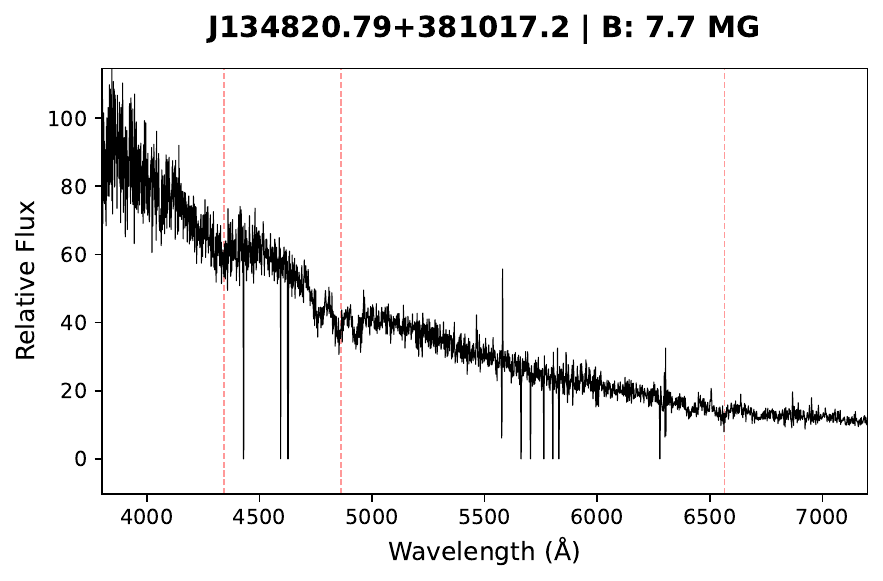}
\includegraphics[width=8.5cm]{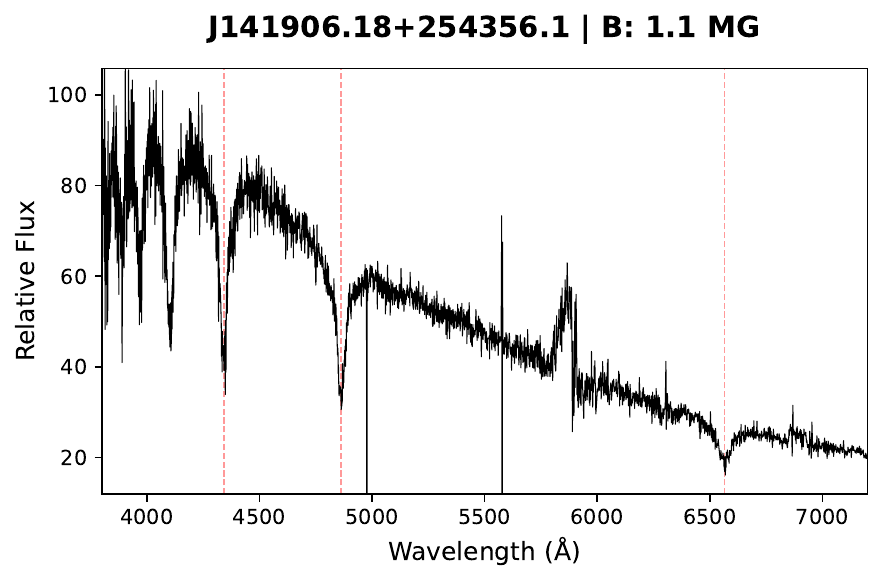}
\includegraphics[width=8.5cm]{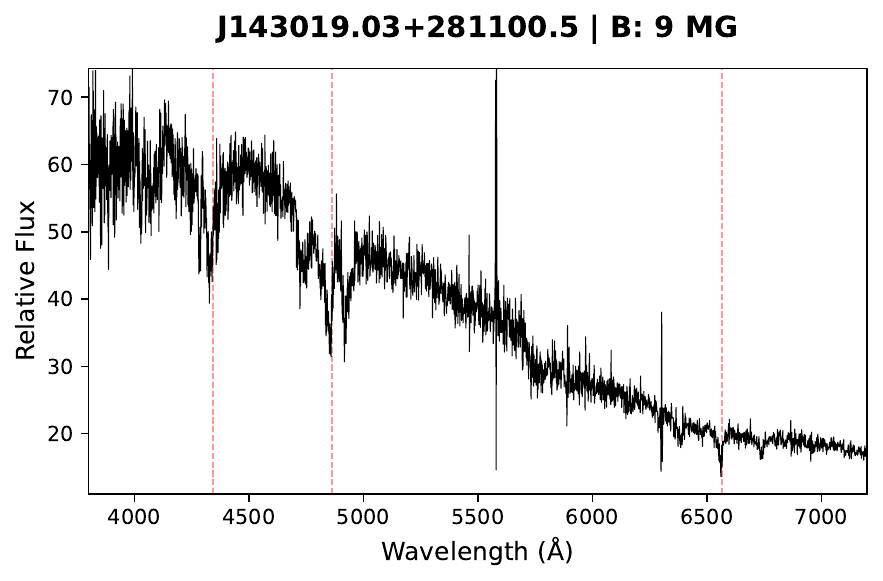}
\includegraphics[width=8.5cm]{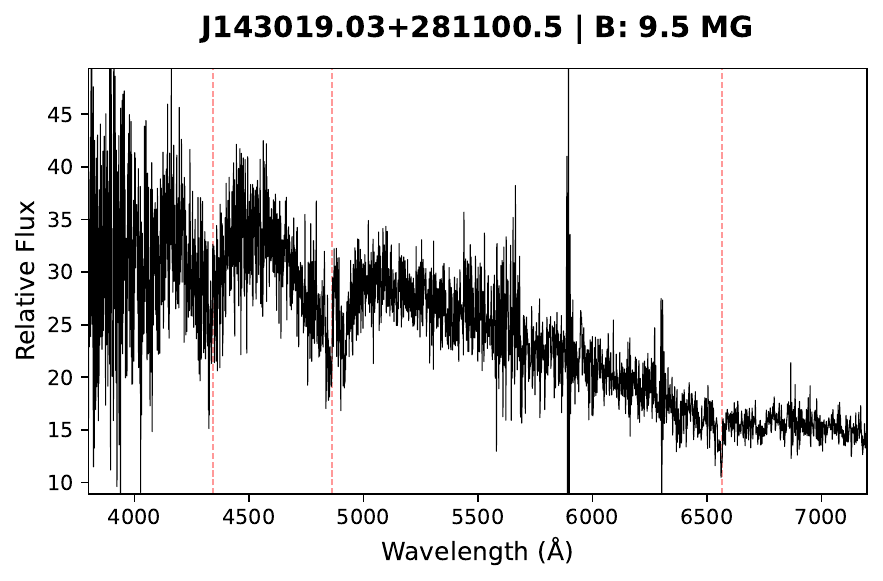}
\includegraphics[width=8.5cm]{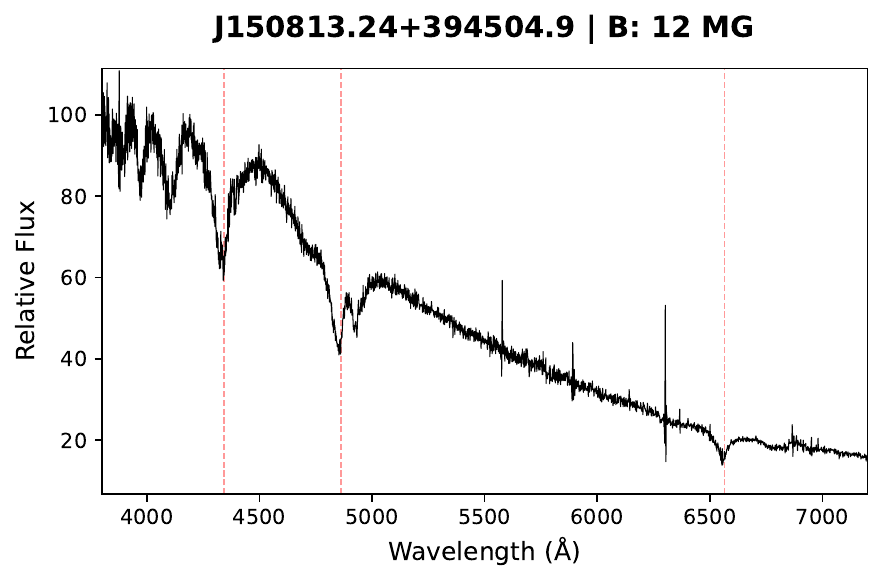}
\includegraphics[width=8.5cm]{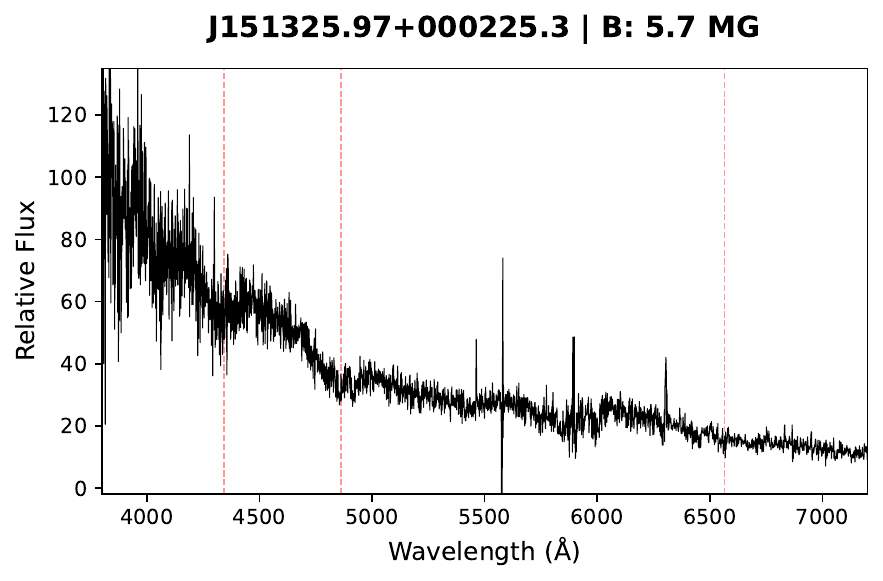}
\includegraphics[width=8.5cm]{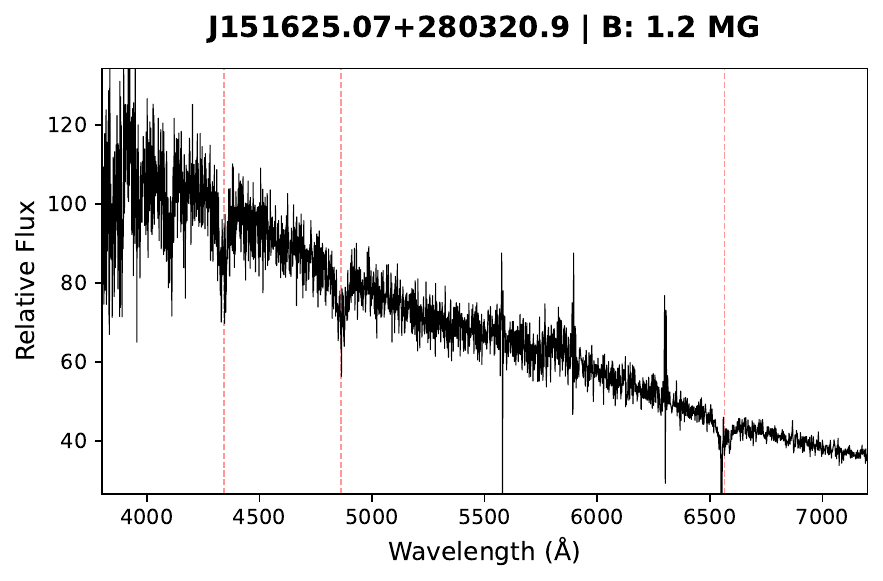}
\includegraphics[width=8.5cm]{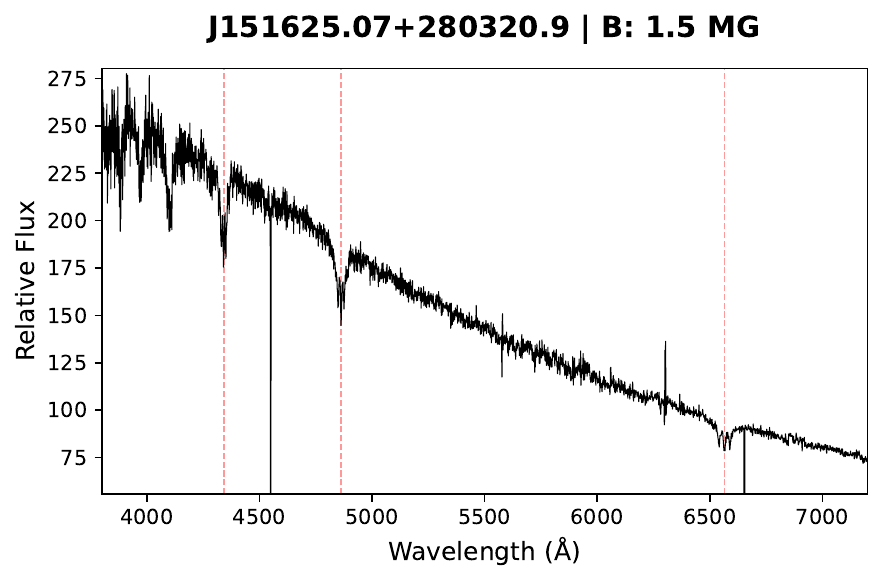}
\end{figure*}

\begin{figure*}
    \centering
\includegraphics[width=8.5cm]{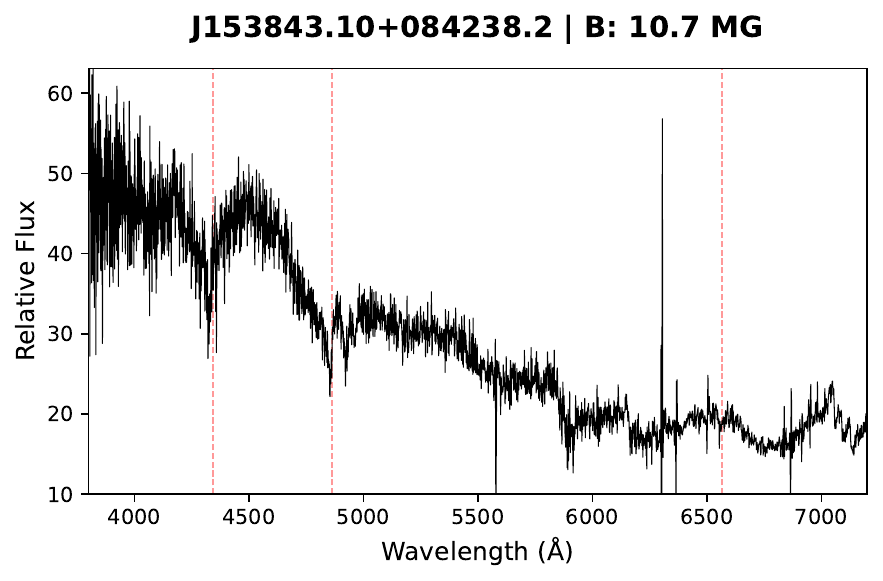}
\includegraphics[width=8.5cm]{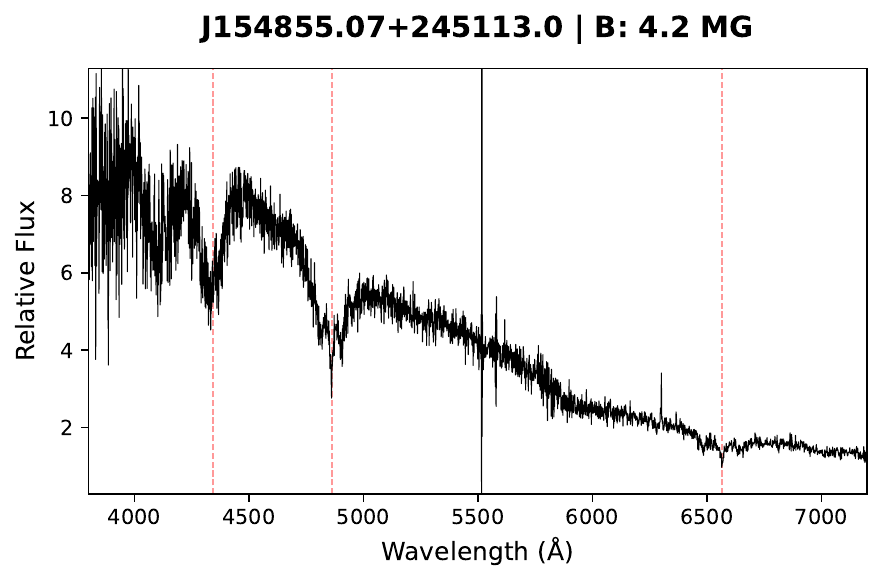}
\includegraphics[width=8.5cm]{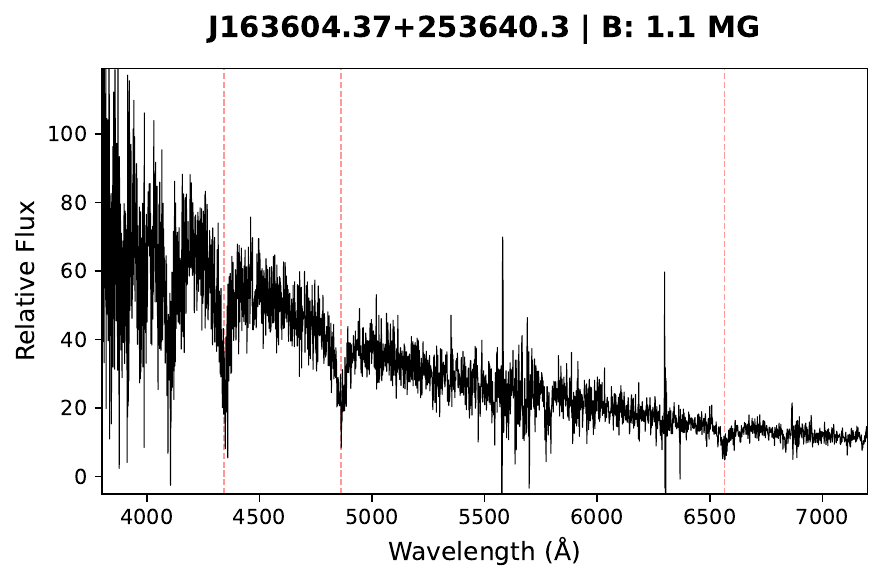}
\includegraphics[width=8.5cm]{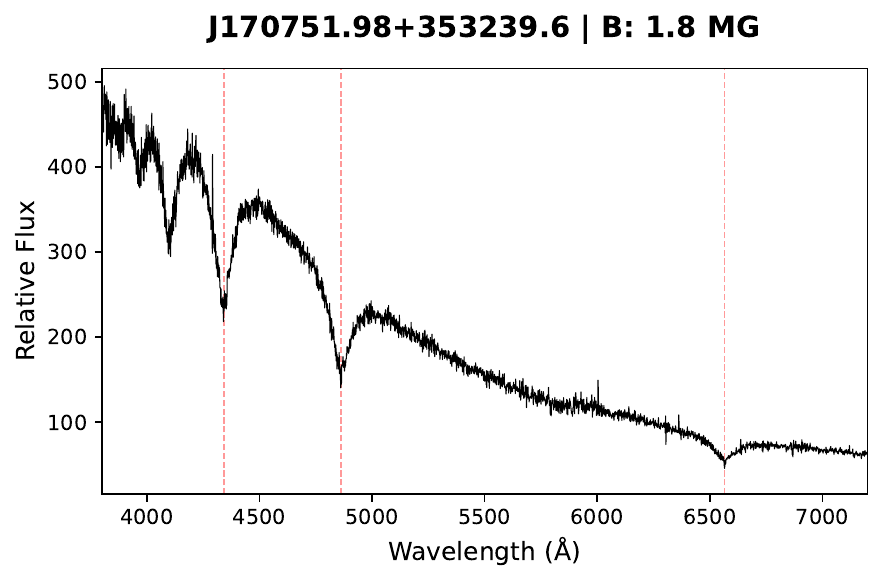}
\includegraphics[width=8.5cm]{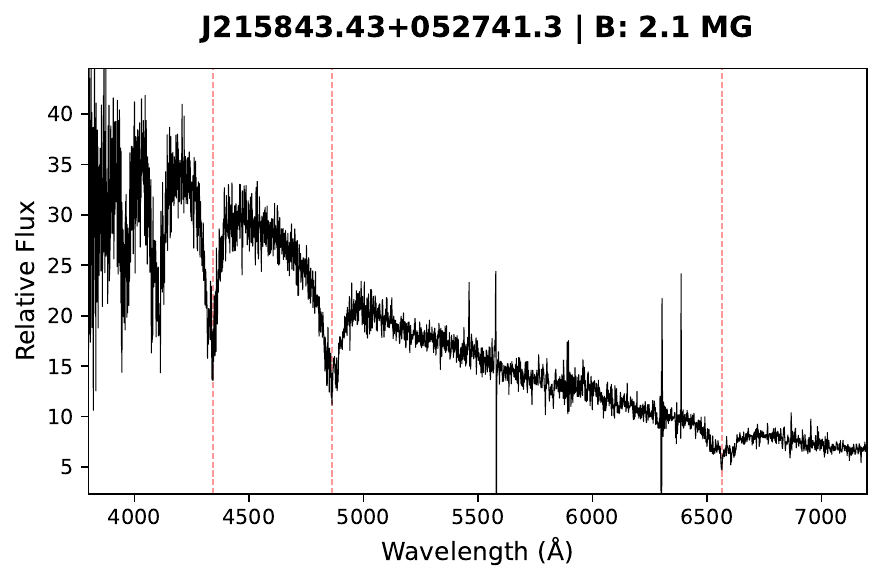}
\includegraphics[width=8.5cm]{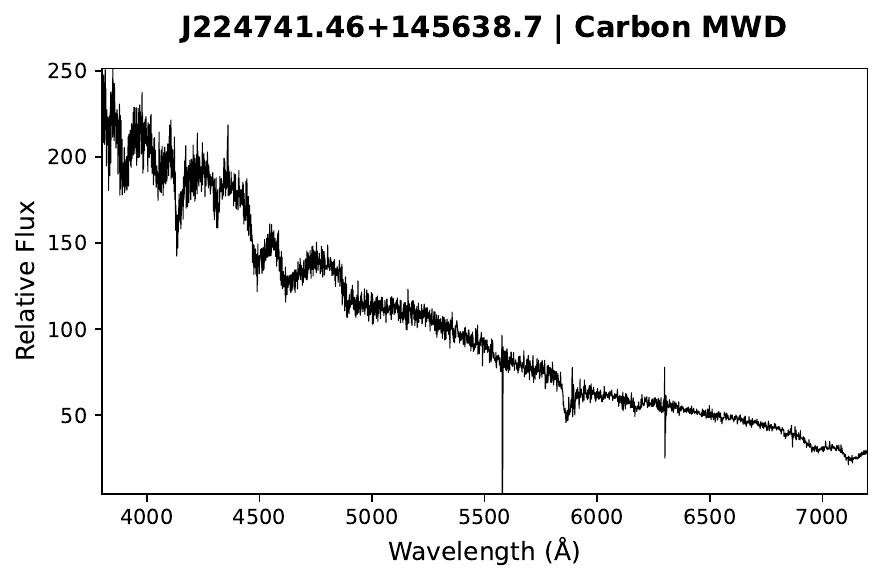}
\includegraphics[width=8.5cm]{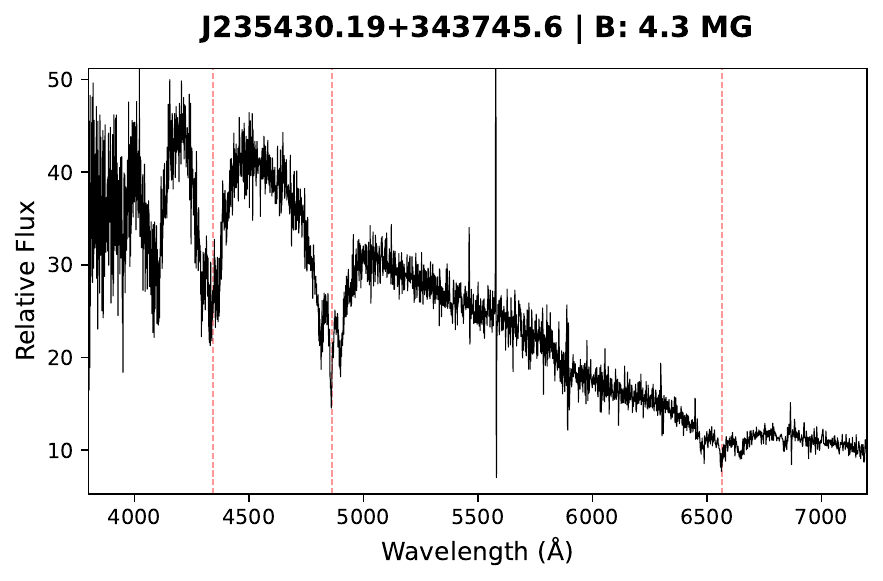}
\includegraphics[width=8.5cm]{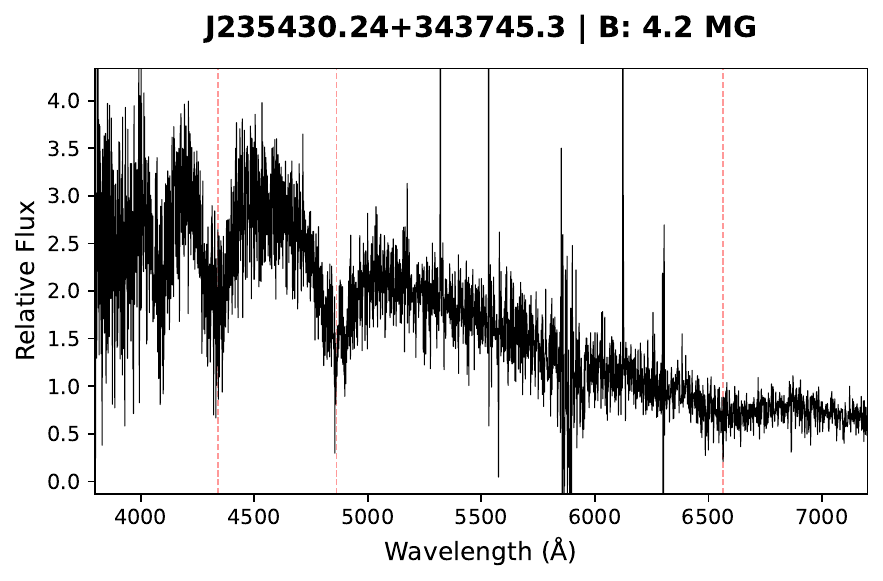}
\end{figure*}


\twocolumn

\end{appendix}

\end{document}